\documentclass[12pt,a4paper]{article}
\usepackage{times}
\usepackage{a4wide}
\usepackage{amsfonts}
\usepackage{amssymb}
\usepackage{amsmath}
\usepackage{ifpdf}
\ifpdf
\usepackage[pdftex,unicode,implicit]{hyperref}
\hypersetup{%
  pdftitle    = {The FGK formalism for  black p-branes in d dimensions},
  pdfkeywords = {supergravity, branes, black branes, attractors},
  pdfauthor   = {Antonio de Antonio Martin, Tomas Ortin, C.S. Shahbazi},
  plainpages  = true,
  colorlinks  = true,
  citecolor   = blue,
  urlcolor    = red,
  linkcolor   = black
}
\newcommand{\hepth}[1]{arXiv:{\tt
\href{http://www.arXiv.org/abs/hep-th/#1}{hep-th/#1}}}

\newcommand{\arxiv}[1]{{\tt
\href{http://www.arXiv.org/abs/#1}{arXiv:#1}}}
\else
  \usepackage[dvips]{graphicx}
  \usepackage[unicode,implicit]{hyperref}
  \newcommand{\hepth}[1]{arXiv:{\tt hep-th/#1}}

  \newcommand{\arxiv}[1]{{\tt arXiv:#1}}
\fi
\makeatletter
\@addtoreset{equation}{section}
\makeatother

\begin{document}

\begin{flushright}
\small
IFT-UAM/CSIC-11-X3\\
\texttt{arXiv:1203.0260 [hep-th]}\\
March  1\textsuperscript{1st}, 2012\\
\normalsize
\end{flushright}

\begin{center}

\vspace{1cm}

{\LARGE {\bf The FGK formalism for black $p$-branes}}\\[8mm]
{\LARGE {\bf in $d$ dimensions}}

\vspace{1.5cm}

\begin{center}

\renewcommand{\thefootnote}{\alph{footnote}}

{\sl\large Antonio de Antonio Mart\'{\i}n}
\footnote{E-mail: {\tt Antonio.de.Antonio.Martin [at] gmail.com}},
{\sl\large Tom\'{a}s Ort\'{\i}n} 
\footnote{E-mail: {\tt Tomas.Ortin [at] csic.es}} 
{\sl\large and C.S.~Shahbazi$^{}$}
\footnote{E-mail: {\tt Carlos.Shabazi [at] uam.es}}

\renewcommand{\thefootnote}{\arabic{footnote}}

\vspace{.5cm}

{\it Instituto de F\'{\i}sica Te\'orica UAM/CSIC\\
C/ Nicol\'as Cabrera, 13--15,  C.U.~Cantoblanco, 28049 Madrid, Spain}\\

\vspace{.5cm}

\end{center}

{\bf Abstract}

\begin{quotation}

  {\small 
    We present a generalization to an arbitrary number of spacetime ($d$) and
    worldvolume ($p+1$) dimensions of the formalism proposed by Ferrara, Gibbons
    and Kallosh to study black holes ($p=0$) in $d=4$ dimensions. We include
    the special cases in which there can be dyonic and self- or anti-self-dual
    black branes. Most of the results valid for 4-dimensional black holes
    (relations between temperature, entropy and non-extremality parameter, and
    between  entropy and black-hole potential on the horizon) are
    straightforwardly generalized. 

    We apply the formalism to the case of black strings in $N=2,d=5$
    supergravity coupled to vector multiplets, in which the black-string
    potential can be expressed in terms of the dual central charge and work
    out an explicit example with one vector multiplet, determining
    supersymmetric and non-supersymmetric attractors and constructing the
    non-extremal black-string solutions that interpolate between them.
}

\end{quotation}

\end{center}

\setcounter{footnote}{0}

\newpage
\pagestyle{plain}

\tableofcontents

\vspace{1cm}

\section*{Introduction and conclusions}

The formalism developed by Ferrara, Gibbons and Kallosh (FGK) in
Ref.~\cite{Ferrara:1997tw} has proven a formidable tool in the study of
4-dimensional black holes. For extremal 4-dimensional black holes, it has
solidly established a connection between the entropy and the values of the
scalars on the horizon through the extremization of the so-called black-hole
potential. In the special case of $N\geq 2,d=4$ supergravity theories, the
black-hole potential is just a function of the central charge and its
covariant derivatives, and some of the extrema of the black-hole potential
(the supersymmetric ones) are the extrema of the central charge, whose value
on the horizon determines the entropy.  This explains how the attractor
mechanism works in these theories \cite{Ferrara:1995ih}

These particular (but very important) results of Ref.~\cite{Ferrara:1997tw}
have been used in much of the literature on black holes: the attractors values
of the scalars on the horizon for a given set of charges of given model or
class of models are determined and the entropy of the corresponding extremal
black holes is computed without ever having to construct the complete
black-hole spacetime metric explicitly. Actually, only in some supergravity
theories it is known how to perform this construction (notably, in $N=2,d=4$
supergravity), even if the generic form of the solutions is known, in
principle, for all 4-dimensional supergravities \cite{Meessen:2010fh}.  For
extremal non-supersymmetric the situation is worse: a systematic procedure to
construct the solutions does not exist even for $N=2,d=4$ supergravities,
except in trivial cases. The non-extremal solutions are described by the FGK
effective action as well and their physics is much richer (and the unknown
extremal non-supersymmetric solutions can probably be be obtained from them);
the absence of an attractor mechanism makes their construction harder and
their study less attractive but definitely no less rewarding.

Recently, a general ansatz to construct general families of non-extremal
black-hole solutions of $N=2,d=4$ supergravity in combination with the FGK
formalism has been proposed in Ref.~\cite{Galli:2011fq} and new variables that
clarify their structure and their construction have been proposed in
Refs.~\cite{Mohaupt:2011aa,Meessen:2011mu}.

Given the power of this approach, it is natural to try to generalize it to
other cases. In Ref.~\cite{Meessen:2011bd} a generalization of the FGK
formalism for $d$-dimensional black holes was presented and the special
properties of the black-hole potential in the $N=2,d=5$ supergravity case were
studied. New variables, similar to those constructed in
Refs.~\cite{Mohaupt:2011aa,Meessen:2011mu} for the 4-dimensional case are also
known \cite{Mohaupt:2009iq,Mohaupt:2010fk,Meessen:2011mu} and all these
results can be combined with the proof of the attractor mechanism presented in
Ref.~\cite{Ortin:2011vm} to find very general results concerning extremal and
non-extremal black-hole solutions of those theories that will be presented
elsewhere \cite{kn:MOPS}.

In this paper we generalize the formalism to $p$-branes in $d$ dimensions,
determining the general form of the metric of a single, charged, static,
regular, flat $p$-brane in $d$ dimensions and constructing the effective
action for the single independent metric function and for the scalars. We
derive the generalization of the results of Ref.~\cite{Ferrara:1997tw} that
relate the values of the scalars on the horizon of extremal black branes to
the extrema of the \textit{black-brane} potential and the entropy to (some
power of) the value of the black-brane potential on the horizon. We also study
the special properties of the black-string potential in the $N=2,d=5$
supergravity case: just as in the black-hole case the black-hole potential
could be written as a function of the central charge and its derivatives, in
the black-string case the black-string potential can be written as a function
of a dual central charge and its derivatives so that the extrema of the
central charge are also (supersymmetric) extrema of the black-string potential
and the entropy is given in those cases by (a power of) the value of the
dual central charge on the horizon. This case is particularly interesting
because new variables, similar to those used for black holes in
Refs.~\cite{Mohaupt:2009iq,Mohaupt:2010fk,Meessen:2011mu} can also be defined
\cite{kn:MOPS}.

Finally, further generalizations to, for instance, branes with curved
worldvolumes such as those considered in Ref.~\cite{Chemissany:2011gr} are
clearly possible using this formalism\footnote{It seems that the background
  transverse metric needs to be modified since, from this point of view, it is
  not universal. We thank T.~Van Riet for pointing out this fact to us.}.

This paper is organized as follows: in Section~\ref{sec:branesformalism} we
describe the general actions we are going to deal with and, using the ansatz
that emerges from Appendix~\ref{sec-known}, we perform the dimensional
reduction to find the generalization of the FGK effective action (obtained in
an alternative fashion in Appendix~\ref{sec-dimred}) and of the general
results concerning extremal branes (Section~\ref{sec:FGKtheorems}).  In
Section~\ref{sec:N2d5} we apply the general formalism to the special case of
black strings in $N=2,d=5$ supergravity and we solve explicitly a simple
model.


\section{The FGK formalism for black $p$- branes}
\label{sec:branesformalism}



\subsection{Derivation of the effective action action}
\label{sec:geodesic action}

We are interested in theories with scalar fields $\phi^{i}$ parametrizing a
non-linear $\sigma$-model with metric $\mathcal{G}_{ij}(\phi)$, and
$(p+1)$-form potentials $A^{\Lambda}_{(p+1)\, \mu_{1}\cdots\mu_{p+2}}$ coupled
to gravity whose actions are of the general form

\begin{equation}
\label{eq:daction}
\mathcal{I}[g,A^{\Lambda}_{(p+1)},\phi^{i}]
=
\int d^{d}x \sqrt{|g|}
\left\{
R + \mathcal{G}_{ij} (\phi)\partial_{\mu} \phi^{i} \partial^{\mu} \phi^{j} 
+4 \tfrac{(-1)^{p}}{(p+2)!} I_{\Lambda \Sigma}(\phi) F_{(p+2)}^{\Lambda} \cdot F_{(p+2)}^{\Sigma}
\right\}\, ,
\end{equation}

\noindent
where 

\begin{equation}
  \begin{array}{rcl}
F^{\Lambda}_{(p+2)\, \mu_{1}\cdots \mu_{p+2}} 
& = & 
(p+2)\partial_{[\mu_{1}|}A^{\Lambda}{}_{(p+1)\, |\mu_{2}\cdots \mu_{p+2}]}\, ,  
\\
& & \\
F_{(p+2)}^{\Lambda} \cdot F_{(p+2)}^{\Sigma}
& \equiv & 
F_{(p+2)\, \mu_{1}\cdots \mu_{p+2}}^{\Lambda}
F_{(p+2)}^{\Sigma}{}^{\mu_{1}\cdots \mu_{p+2}}\, ,
\end{array}
\end{equation}

\noindent
are the $(p+2)$-form field strengths and $I_{\Lambda \Sigma}(\phi)$ is a
scalar-dependent, negative-definite matrix that describes the coupling of the
scalar fields to the $(p+1)$-form fields. The normalizations have been chosen
so as to recover the particular cases considered in
Refs.~\cite{Meessen:2011bd,Ferrara:1997tw} for $p=0$, general $d$ and
$p=0,d=4$, respectively, with the original normalizations.

In the particular cases $p=\tilde{p}=(d-4)/2$ (for instance, black holes in
$d=4$, strings in $d=6$, membranes in $d=8$ and 3-branes in $d=10$, to mention
only those which are relevant from the String Theory point of view) one should
consider additional terms of the form

\begin{equation}
  \label{eq:additionalterm}
+4\xi^{2} \tfrac{(-1)^{p}}{(p+2)!}R_{\Lambda\Sigma}(\phi) 
F_{(p+2)}^{\Lambda} \cdot \star F_{(p+2)}^{\Sigma}\, ,  
\end{equation}

\noindent
in the action, where $R_{\Lambda\Sigma}(\phi)$ is a scalar dependent matrix
such that

\begin{equation}
R_{\Lambda\Sigma} = -\xi^{2}R_{\Sigma\Lambda}\, ,  
\end{equation}

\noindent
and where\footnote{This constant is associated to the value of the square of
  the Hodge star when it acts on a $(p+2)$ form: $\star^{2} = \xi^{2}$.}

\begin{equation}
\xi^{2} =-(-1)^{d/2} = (-1)^{p+1}\, ,
\end{equation}

\noindent
and the ansatz should take into account that the same brane can also be
magnetically charged with respect to the dual of the $(p+1)$-form potentials,
which are also $(p+1)$-forms, i.e.~they can be dyonic. Furthermore, if $d=
4n+2$ ($p$ odd: strings in $d=6$ and 3-branes in $d=10$) the dyonic branes can
also be self- or anti-self-dual. 

The first ingredient we need is a generic ansatz for the metric of any
electrically charged, static, flat, black $p$-brane in $d=p+\tilde{p}+4$
dimensions,  where $\tilde{p}$ is the dimension of the of the dual (magnetic)
brane, with a transverse radial coordinate $\rho$ such that the event horizon
is at $\rho\rightarrow \infty$.

This generic ansatz can be found by studying the metrics of known families of
solutions of this kind, such as those originally found in
Ref.~\cite{Horowitz:1991cd}\footnote{Here we use the conventions and notation
  of Ref.~\cite{Ortin:2004ms}}. This study is performed in
Appendix~\ref{sec-known} and the ansatz for the metric that emerges from it
is\footnote{This metric has also been derived from the equatios of motion in
  Refs.~\cite{Janssen:2007rc}, where it has also been shown to be valid for
  time-dependent cases. Inthose references, more general slicings of the
  spacetime were also considered.}

\begin{equation}
\label{eq:generalmetric1}
ds_{(d)}^{2}
= 
e^{\frac{2}{p+1}\tilde{U}}
\left[
W^{\frac{p}{p+1}} dt^{2}
-W^{-\frac{1}{p+1}}d\vec{y}^{\, 2}_{(p)}
\right]  
-e^{-\frac{2}{\tilde{p}+1}\tilde{U}}
\gamma_{(\tilde{p}+3)\, \underline{m}\underline{n}}  dx^{m} dx^{n}\, .
\end{equation}

\noindent
where $\vec{y}_{(p)}\equiv (y^{1},\cdots, y^{p})$ are the brane's $p$
spacelike worldvolume coordinates and where $\gamma_{(\tilde{p}+3)\,
  \underline{m}\underline{n}}$ is the background transverse metric given by

\begin{equation}
\label{eq:backgroundtransversemetric}
\gamma_{(\tilde{p}+3)\, \underline{m}\underline{n}}  dx^{m} dx^{n}
= 
\left(\frac{ \omega/2}{\sinh{\left(\frac{\omega}{2} \rho\right)}} \right)^{\frac{2}{\tilde{p}+1}}
\left[ 
\left( \frac{\omega/2}{\sinh{\left(\frac{\omega}{2} \rho \right)}}\right)^2
\frac{d\rho^2}{(\tilde{p}+1)^2}
+ d\Omega^{2}_{(\tilde{p}+2)} 
\right]\, ,	 
\end{equation}

\noindent
where, in turn, $d\Omega_{(\tilde{p} +2)}^{2}$ is the metric of the
round $(\tilde{p}+2)$-sphere of unit radius.

The general metric Eq.~(\ref{eq:generalmetric1}), which reduces in the $p=0$
case to the metrics used in $d=4$ and arbitrary $d$-dimensional black holes in
Refs.~\cite{Ferrara:1997tw} and \cite{Meessen:2011bd} respectively ($W$
disappears), should be capable of describing any non-extremal black brane for
adequate choices of the functions $\tilde{U}(\rho)$ and $W(\rho)$. In what
follows we will use it as an ansatz in which only $\tilde{U}(\rho)$ and
$W(\rho)$ have to be determined.

Observe that, while it is possible to redefine $\tilde{U}$ and the transverse
metric $\gamma_{(\tilde{p}+3)\, \underline{m}\underline{n}}$ so as to totally
absorb $W$ in some components of the metric, it is not possible to do it
simultaneously in all of them.  We do not expect more than one independent
function in a black-brane metric, but nothing prevents us from using the above
metric with \textit{a priori} independent functions $\tilde{U}$ and $W$ as
an ansatz and then letting the equations of motion dictate how they are related
and what is the best definition for the single independent function that we
expect.

If we are to describe electrically charged $p$-branes, an adequate ansatz for
the $(p+1)$-form potentials $A^{\Lambda}_{(p+1)}$ is

\begin{equation}
A^{\Lambda}_{(p+1)\, ty_{1}\cdots y_{p}} = \psi^{\Lambda}(\rho)\, ,  
\end{equation}

\noindent
(all the other components vanish). In the special case $p=\tilde{p}=(d-4)/2$,
the branes can also be magnetically charged with respect to the dual
(\textit{magnetic}) $(p+1)$-form potentials. These are defined as follows: the
equations of motion of the \textit{electric} $(p+1)$-form potentials, when we
add the term Eq.~(\ref{eq:additionalterm}) to the action, can be expressed in
the form

\begin{equation}
dG_{(p+2)\, \Lambda}=0\, ,
\hspace{1cm}
G_{(p+2)\,\Lambda} \equiv 
R_{\Lambda\Sigma}F_{(p+2)}^{\Sigma}+I_{\Lambda\Sigma}\star F_{(p+2)}^{\Sigma}\, ,  
\end{equation}

\noindent
and imply the local existence of the magnetic $(p+1)$-form potentials
$A_{(p+1)\,\Lambda}$ satisfying

\begin{equation}
G_{(p+2)\,\Lambda} = d A_{(p+1)\,\Lambda}\, .
\end{equation}

\noindent
Then, in this particular cases, our ansatz for the magnetic potentials is 

\begin{equation}
A_{(p+1)\, \Lambda\,  ty_{1}\cdots y_{p}} = \chi_{\Lambda}(\rho)\, .  
\end{equation}

The electric and magnetic field $(p+2)$-form strengths can be arranged into a
vector

\begin{equation}
\left(\mathcal{F}^{M}\right) 
\equiv  
\left(
\begin{array}{c}
F^{\Lambda} \\ G_{\Lambda} \\
\end{array}
\right)\, ,
\hspace{1cm}
\left(\Psi^{M}\right) 
\equiv  
\left(
\begin{array}{c}
\psi^{\Lambda} \\ \chi_{\Lambda} \\
\end{array}
\right)\, ,
\end{equation}

\noindent
so the Bianchi identities and Maxwell equations can be written in the compact
form

\begin{equation}
d\mathcal{F}^{M}=0\, ,  
\end{equation}

\noindent
which is covariant under linear transformations

\begin{equation}
\left(
\begin{array}{c}
F^{\prime} \\ G^{\prime} \\
\end{array}
\right)
= 
\left(
\begin{array}{cc}
A & B \\
C & D \\    
\end{array}
\right)
\left(
\begin{array}{c}
  F \\ G \\
\end{array}
\right)\, .  
\end{equation}

\noindent
The consistency of these linear transformations with the definitions of the
magnetic field strengths requires that the matrices $R,I$ transform according
to

\begin{equation}
N^{\prime}
=
\left(C+D N\right)
\left(A+BN\right)^{-1}\, ,
\hspace{1cm}
N \equiv R+\xi I\, .
\end{equation}

On the other hand, the contribution of the $(p+1)$-form potentials to the
energy-momentum tensor can be written in the form

\begin{equation}
\Omega_{MN} \star\mathcal{F}^{M}{}_{\mu\alpha_{1}\cdots \alpha_{p+1}}  
\mathcal{F}^{N}{}_{\nu}{}^{\alpha_{1}\cdots \alpha_{p+1}}\, ,
\end{equation}

\noindent
where we have defined the metric

\begin{equation}
\left(\Omega_{MN} \right)
\equiv 
\left(
\begin{array}{cc}
0  & \mathbb{I} \\
\xi^{2}\mathbb{I} & 0 \\   
\end{array}
\right)\, ,
\end{equation}

\noindent
which will be used to raise and lower $M,N$ indices.  This implies that the
linear transformations of the $n$ electric and $n$ magnetic field strengths
must be restricted to $O(n,n)$ when $\xi^{2}=+1$ and to $Sp(2n+2,\mathbb{R})$ when
$\xi^{2}=-1$.

An alternative expression for this contribution to the energy-momentum tensor
is

\begin{equation}
\mathcal{M}_{MN}\mathcal{F}^{M}{}_{\mu\alpha_{1}\cdots \alpha_{p+1}}  
\mathcal{F}^{N}{}_{\nu}{}^{\alpha_{1}\cdots \alpha_{p+1}}\, ,  
\end{equation}

\noindent
where the symmetric matrix $\mathcal{M}_{MN}$ is given by 

\begin{equation}
  \begin{array}{rcl}
\left(\mathcal{M}_{MN} \right)
& \equiv &
\left(
\begin{array}{cc}
I-\xi^{2}RI^{-1}R  & \xi^{2} RI^{-1} \\
& \\
-I^{-1}R & I^{-1} \\   
\end{array}
\right)\, ,
\\
& & \\
\left(\mathcal{M}^{MN} \right)
& = &
\left(
\begin{array}{cc}
I^{-1} & -\xi^{2} I^{-1}R \\   
& \\
 RI^{-1} & I-\xi^{2}RI^{-1}R  \\
\end{array}
\right)=
(\mathcal{M}_{NP})^{-1}\, .
\end{array}
\end{equation}

In what follows we will write the expressions including the additional terms
(matrix $R_{\Lambda\Sigma}$, magnetic charges $p^{\Lambda}$ etc.) in the
understanding that they vanish whenever the condition $p=\tilde{p}=(d-4)/2$ is
satisfied.

To end the description of our ansatz, we are also going to assume that the
scalars only depend on $\rho$.  Plugging this ansatz into the equations of
motion derived from the above action, we get two equations

\begin{eqnarray}
\frac{d^{2}\ln{W}}{d\rho^{2}}
& = & 
0\, ,
\\
& & \nonumber \\
\frac{d~}{d\rho}\left[e^{-2\tilde{U}}\, \mathcal{M}_{MN}\, \dot{\Psi}^{N} \right]
& = & 
0\, .
\end{eqnarray}

\noindent
(overdots denoting derivatives w.r.t.~$\rho$)
that can be integrated immediately, giving

\begin{eqnarray}
W
& = & 
e^{\gamma \rho}\, ,
\\
& & \nonumber \\
\dot{\Psi}^{M}  
& = & 
\alpha e^{2\tilde{U}}\mathcal{M}^{MN}\mathcal{Q}_{N}\, ,
\end{eqnarray}

\noindent
where we have normalized $W(0)=1$ at spatial infinity and we have introduced
the integration constants $\gamma$ and $\mathcal{Q}_{M}$, $\alpha$ being a
normalization constant. The constants $\mathcal{Q}_{M}$ are, up to global
normalization, just the electric and magnetic charges of the $p$-brane with
respect to the $(p+1)$-form potentials

\begin{equation}
  \mathcal{Q}_{M} \sim \int_{S^{\tilde{p}+2}} \star\, \mathcal{M}_{MN}
  \mathcal{F}^{N}\, ,
\hspace{1cm}
(\mathcal{Q}^{M}) \equiv 
\left(
\begin{array}{c}
p^{\Lambda} \\ q_{\Lambda} \\    
\end{array}
\right)\, ,
\hspace{1cm}
\mathcal{Q}_{M} \equiv \Omega_{MN}\mathcal{Q}^{N}\, .
\end{equation}

These first integrals allow us to eliminate from the equations of motion $W$
and $\Psi^{M}$ (which only appears through $\dot{\Psi}^{M}$). The remaining
three equations only involve $\tilde{U}$ and $\phi^{i}$ and take the form

\begin{eqnarray}
\label{eq:1}
\ddot{\tilde{U}} +e^{2\tilde{U}}V_{\rm BB}
& = & 
0\, ,
\\
& & \nonumber \\
\ddot{\phi}^{i} +\Gamma_{jk}{}^{i}\dot{\phi}^{j}\dot{\phi}^{k}
+\tfrac{d-2}{2(\tilde{p}+1)(p+1)}e^{2\tilde{U}}\partial^{i} V_{\rm BB}
& = & 
0\, ,  
\\
& & \nonumber \\
\label{eq:hamiltonianconstraint}
(\dot{\tilde{U}})^{2} 
+\tfrac{(p+1)(\tilde{p}+1)}{d-2} \mathcal{G}_{ij} \dot{\phi}^{i} \dot{\phi}^{j}
+e^{2 \tilde{U}} V_{\rm BB}
&  = & 
\hat{\mathcal{B}}^{2}\, ,
\end{eqnarray}

\noindent
where we have defined the negative semidefinite \textit{black-brane potential}

\begin{equation}
\label{eq:VBB}
V_{\rm BB}(\phi,\mathcal{Q}) \equiv 
2\alpha^{2} \tfrac{(p+1)(\tilde{p}+1)}{(d-2)} 
\mathcal{M}_{MN} \mathcal{Q}^{M}\mathcal{Q}^{N}\, ,
\end{equation}

\noindent
and the constant 

\begin{equation}
\label{eq:B2}
\hat{\mathcal{B}}^{2} \equiv \tfrac{(p+1)(\tilde{p}+2)}{4(d-2)}\, \omega^{2} 
-\tfrac{(\tilde{p}+1)p}{4(d-2)}\gamma^{2}\, .
\end{equation}

These equations (up to the constant in Eq.~(\ref{eq:hamiltonianconstraint},
which arises as the Hamiltonian constraint) can be derived from from the
effective action

\begin{equation}
\label{eq:geodesicaction}
\mathcal{I}[\tilde{U},\phi^{i}]=
\int d\rho 
\left\{
(\dot{\tilde{U}})^{2}
+\tfrac{(p+1)(\tilde{p}+1)}{d-2} \mathcal{G}_{ij} \dot{\phi}^{i} \dot{\phi}^{j}
-e^{2 \tilde{U}} V_{\rm BB} +\hat{\mathcal{B}}^{2}
\right\}\, .
\end{equation}
 
Summarizing, we have found that, if we use the ansatz 

\begin{equation}
\begin{array}{rcl}
ds_{(d)}^{2}
& = &
e^{\frac{2}{p+1}\tilde{U}}
\left[
e^{\frac{p}{p+1}\gamma\rho} dt^{2}
-e^{-\frac{1}{p+1}\gamma\rho}d\vec{y}^{\, 2}_{(p)}
\right]  
-
e^{-\frac{2}{\tilde{p}+1}\tilde{U}}
\gamma_{(\tilde{p}+3)\, \underline{m}\underline{n}}  dx^{m} dx^{n}\, ,
\\
& & \\
A^{M}_{(p+1)} 
& = & 
\Psi^{M}(\rho)\, dt\wedge dy^{1} \wedge \cdots \wedge dy^{p}\, ,  
\hspace{1cm}
\dot{\Psi}^{M }  
=
\alpha e^{2\tilde{U}}\mathcal{M}^{MN}\mathcal{Q}_{N}\, ,
\\
& & \\
\phi^{i}
& = & 
\phi^{i}(\rho)\, ,
\end{array}
\end{equation}

\noindent
where $\tilde{U}$ is a function of $\rho$; $\gamma,\mathcal{Q}_{M}$ are constants
and $\gamma_{(\tilde{p}+3)\, \underline{m}\underline{n}}$ is the transverse
space metric given in Eq.~(\ref{eq:generalmetric1}), in the theories defined by
generic family of actions Eq.~(\ref{eq:daction}), we find that they are
solutions of these theories if the following
Eqs.~(\ref{eq:1})-(\ref{eq:hamiltonianconstraint}) are
satisfied.

The same result is obtained in Appendix~\ref{sec-dimred} by reducing first the
action Eq.~(\ref{eq:daction}) to $(d-p)=(\tilde{p}+4)$ dimensions  in
such a way that the action only contains the Einstein-Hilbert term, scalars
and 1-forms and then by using the FGK formalism of Ref.~\cite{Meessen:2011bd}
in a second stage.

In general, the integration constant $\gamma$ will be related to the
non-extremality parameter $\omega$ by requiring the solution to have a regular
event horizon. Indeed, Eq.~(\ref{eq:asympW}) implies that  

\begin{equation}
\gamma=\omega
\hspace{1cm}
W=e^{\omega\rho}\, ,
\hspace{1cm}
\hat{\cal B}^{2}= (\omega/2)^{2}\, ,  
\end{equation}

\noindent
and, therefore, the general form of regular $p$-branes will be taken to be

\begin{equation}
\label{eq:generalmetric}
ds_{(d)}^{2}
= 
e^{\frac{2}{p+1}\tilde{U}}
\left[
e^{\frac{p}{p+1}\omega\rho} dt^{2}
-e^{-\frac{1}{p+1}\omega\rho}d\vec{y}^{\, 2}_{(p)}
\right]  
-
e^{-\frac{2}{\tilde{p}+1}\tilde{U}}
\gamma_{(\tilde{p}+3)\, \underline{m}\underline{n}}  dx^{m} dx^{n}\, ,  
\end{equation}


\subsection{FGK theorems for static flat branes}
\label{sec:FGKtheorems}


In the same spirit as \cite{Ferrara:1997tw,Meessen:2011bd}, we can use the
formalism presented in the previous section to derive several results about
single, static, flat, black $p$-brane solutions in $d$ dimensions. 

Let us first consider extremal black branes $\omega=0$, whose general form
follows from the $\omega\longrightarrow 0$ limit of the general metric
Eq.~(\ref{eq:generalmetric}):

\begin{equation}
\label{eq:extremalgeneralmetric}
ds_{(d)}^{2}  
=  
e^{\frac{2\tilde{U}}{p+1}} \left[dt^{2}- d\vec{y}_{(p)}^{~2} \right]      
-\frac{e^{-\frac{2\tilde{U}}{\tilde{p}+1}}}{\rho^{\frac{2}{\tilde{p}+1}}} 
\left[ \frac{1}{\rho^{2}}\frac{d\rho^{2}}{(\tilde{p}+1)^2}
+d\Omega^{2}_{(\tilde{p}+2)}\right] \, .
\end{equation}

\noindent
According to the results in Appendix~\ref{sec-extremal}, in the extremal
limit, $\tilde{U}$ must behave as in Eq.~(\ref{eq:Uasymp}), which
we reproduce here for convenience:

\begin{equation}
e^{\tilde{U}}
\sim
\tilde{S}^{-\frac{\tilde{p}+1}{\tilde{p}+2}} \rho^{-1}\, ,
\end{equation}

\noindent
where $\tilde{S}$ is the entropy density per unit worldvolume , defined in the
paragraph above Eq.~(\ref{eq:entropydensity}).  Therefore, the near-horizon
limit of Eq.~(\ref{eq:extremalgeneralmetric}) takes the general form

\begin{equation}
\label{eq:metricextremalimit}
ds^{2}_{(d)}  
= 
\rho^{\frac{-2}{p+1}} \tilde{S}^{-\frac{2(\tilde{p}+1)}{(p+1) (\tilde{p}+2)}} 
\left[dt^2- d\vec{y}_{(p)}^{~2} \right]      
-\tilde{S}^{\frac{2}{\tilde{p}+2}}\left[
  \frac{1}{\rho^2}\frac{d\rho^2}{(\tilde{p}+1)^2}
+d\Omega^{2}_{(\tilde{p}+2)}\right] \, ,
\end{equation}

\noindent
which is the direct product $AdS_{p+2}\times S^{\tilde{p}+2}$, both with radii
dual to $\tilde{S}^{\frac{1}{\tilde{p}+2}}$.

We impose the following regularity condition on the scalars

\begin{equation}
\label{eq:scalarregularity}
\lim_{\rho\rightarrow \infty} \frac{(p+1)(\tilde{p}+1)}{d-2} 
\mathcal{G}_{ij}\dot{\phi}^i  \dot{\phi}^j 
 e^{2\tilde{U}} \rho^{4}\equiv \mathcal{X} < \infty\, .
\end{equation}

\noindent
Then, the near-horizon limit $\rho\rightarrow\infty$ of the Hamiltonian
constraint Eq.~(\ref{eq:hamiltonianconstraint}) is

\begin{equation}
\label{eq:nearhorizonantigravity}
1+\mathcal{X} \tilde{S}^{\frac{2(\tilde{p}+1)}{\tilde{p}+2}}
+\tilde{S}^{-\frac{\tilde{p}+1}{\tilde{p}+2}} V_{\rm BB}(\phi_{H},\mathcal{Q})=0\, .
\end{equation}

\noindent
If we assume that the entropy density $\tilde{S}$ does not vanish and the
values of the scalars do not diverge on the horizon $\phi^{i}_{\rm h}<
\infty$, then it can be shown that 

\begin{equation}
\label{eq:deduccion1}
\rho\frac{d\phi^i}{d\rho}=0,~~~ \mathcal{X}=0\, ,
\end{equation}

\noindent
and from Eqs.~(\ref{eq:nearhorizonantigravity}) and (\ref{eq:deduccion1}) we
obtain

\begin{equation}
\label{eq:extremalarea}
\tilde{S}
= 
\left[-V_{\rm BB}(\phi_{\rm h},\mathcal{Q})
\right]^{\frac{\tilde{p}+2}{2(\tilde{p}+1)}}\, ,
\end{equation}

\noindent
and therefore the entropy of an extremal brane is given by (a power of) the
value of the black-brane potential at the horizon. 

On the other hand, if we assume that, again, the entropy density is finite
and, furthermore, that

\begin{equation}
\label{eq:assume2}
\rho\frac{d\phi^i}{d\rho}=0\,  ,\,\,\,\,\ \forall i\, ,
\end{equation}

\noindent
we deduce, from the near-horizon limit of the equations of the scalars, that
the value of the scalars on the horizon is fixed in terms of the charges by 

\begin{equation}
\label{eq:attractor1}
\mathcal{G}^{ij}(\phi_{\rm h})\partial_i V_{\rm BB}(\phi_{\rm
  h},\mathcal{Q})=0\, ,
\end{equation}

\noindent
and does not diverge.

Therefore the condition Eq.~(\ref{eq:assume2}) plus finiteness of the entropy
density imply the regularity of the scalars on the horizon that we assumed
before. If the metric of the scalar manifold $\mathcal{G}_{ij}$ is positive
definite, then Eq.~(\ref{eq:attractor1}) is equivalent to

\begin{equation}
\label{eq:attractor2}
\partial_{i} V_{\rm BB}(\phi_{\rm h},\mathcal{Q})=0\, ,
\end{equation}

\noindent
which generalizes the usual attractor mechanism for static extremal black
holes to the case of static extremal flat branes.

Finally, if we take the spatial infinity limit $\rho\rightarrow 0^{+}$ of the
Hamiltonian constraint Eq.~(\ref{eq:hamiltonianconstraint}), we obtain the
analog for branes of the so-called extremality (or antigravity) bound for
black holes

\begin{equation}
\label{eq:antigravitybound}
\tilde{u}^2
+\frac{(p+1) (\tilde{p}+1)}{d-2} \mathcal{G}_{ij}(\phi_{\infty})\Sigma^i
\Sigma^j 
+V_{BB} (\phi_{\infty},\mathcal{Q})=(\omega/2)^2 \, ,
\end{equation}

\noindent
where $\Sigma^i$ are the scalar charges and $\tilde{u}=-\tilde{U}'(0)$ is 
given in terms of the black $p$-brane's tension $T_{p}$ and the
non-extremality parameter $\omega$ by Eq.~(\ref{eq:generaltensionformula}):

\begin{equation}
\tilde{u} = -\frac{1}{(d-2)}
\left[
(p+1)(\tilde{p}+2)T_{p} +p(\tilde{p}+1)\omega/2
\right]\, .  
\end{equation}

The above formula differs from the black hole's by terms proportional to
$p\omega$ which vanish in the black-hole case $p=0$.


\section{Non-extremal strings in $N=2$, $d=5$ supergravity.}
\label{sec:N2d5}


In order to illustrate the formalism developed in the previous sections, we
are going to particularize it for the case of $N=2,d=5$ supergravity, solving
a simple example. The relevant part of the bosonic action of $N=2,d=5$
supergravity theories coupled to $n$ vector multiplets is, using the
conventions of Refs.~\cite{Bellorin:2006yr,Bergshoeff:2004kh},

\begin{equation}
\label{eq:N2d5action}
\mathcal{I}[g_{\mu\nu},A^{I}{}_{\mu},\phi^{x}] = \int d^{5}x 
\left\{
R  +\tfrac{1}{2}g_{xy}\partial_{\mu}\phi^{x}\partial^{\mu}\phi^{y} 
-\tfrac{1}{4}a_{IJ}F^{I}{}_{\mu\nu}F^{J\, \mu\nu}
\right\}\, ,
\end{equation}

\noindent
where $I,J=0,1,\cdots,n$ and $x,y=1,\cdots,n$. The scalar target spaces are
implicitly defined by the existence of $n+1$ functions $h^{I}(\phi)$ of the
$n$ physical scalar subject to the constraint

\begin{equation}
\label{eq:hypersurface}
C_{IJK}h^{I}h^{J}h^{K}=1\, ,  
\end{equation}

\noindent
where $C_{IJK}$ is a completely symmetric constant tensor that determines the
model. Defining

\begin{equation}
  h_{I} \equiv C_{IJK}h^{J}h^{K}\, ,\hspace{1cm}\mbox{(so $h_{I}h^{I}\ =\
    1$)}\, ,
\end{equation}

\noindent
the positive definite matrix $a_{IJ}$ can be expressed as

\begin{equation}
a_{IJ} = -2C_{IJK}h^{K}+3h_{I}h_{J}\, ,  
\end{equation}

\noindent
and can be used to consistently raise and lower the index of the functions
$h^{I}$. We also define

\begin{equation}
  \label{eq:9}
  h^{I}{}_{x} \ \equiv\ -\sqrt{3} \partial_{x}h^{I}\;\; ,\;\;
  h_{I\, x} \ \equiv\ a_{IJ}h^{J} \ =\ +\sqrt{3} \partial_{x}h_{I}\; ,
\end{equation}

\noindent
which are orthogonal to the functions $h^{I}$ with respect to the metric
$a_{IJ}$. Finally, the $\sigma$-model metric is given by

\begin{equation}
\label{eq:a-1}
g_{xy}\  \equiv\  a_{IJ}h^{I}{}_{x} h^{J}{}_{y} \;\;
\longrightarrow
\;\;
a^{IJ} = h^{I}h^{J} +g^{xy}h^{I}{}_{x}h^{J}{}_{y}\, ,
\hspace{.5cm}
a^{IJ}a_{JK}= \delta^{I}{}_{K}\, .
\end{equation}

Since we want to obtain non-extremal strings, it is more convenient to use the
dual 2-form potentials $B_{I\, \mu\nu}$ and their 3-form field strengths
$H_{I\, \mu\nu\rho}= 3\partial_{[\mu|}B_{I\, |\nu\rho]}$, which are related to
the 1-forms by the duality relations

\begin{equation}
\label{eq:dualvariables}
H_{I} = a_{IJ}\star F^{J}\, .
\end{equation}

In terms of these variables the action takes the form\footnote{We have
  dualized an incomplete action Eq.~(\ref{eq:N2d5action}), and, therefore,
  there are terms missing in this dual action. However, for the kind of
  solutions that we want to study, only electrical charged with respect to
  the 2-forms, the missing terms are irrelevant.}

\begin{equation}
\label{eq:actionsimplemodel}
\mathcal{I}[g_{\mu\nu},B^{I}{}_{\mu\nu},\phi^{x}] = \int d^{5}x 
\left\{
R  +\tfrac{1}{2}g_{xy}\partial_{\mu}\phi^{x}\partial^{\mu}\phi^{y} 
+\tfrac{1}{2\cdot 3!}a^{IJ}H_{I}{}^{\mu\nu\rho}H_{J\, \mu\nu\rho}
\right\}\, ,
\end{equation}

Comparing now Eq.~(\ref{eq:actionsimplemodel}) (taking $p=1$, $\tilde{p}=0$,
as corresponds to $d=5$ string solutions) to Eq.~(\ref{eq:daction}) we find
that

\begin{equation}
\label{eq:relaciones}
I_{IJ}= -\tfrac{1}{8} a^{IJ}\, ,
\hspace{1cm}
\mathcal{G}_{xy}=\tfrac{1}{2} g_{xy}\, ,
\end{equation}

\noindent
and, therefore, the effective action for this model is given by

\begin{equation}
\label{eq:effectived5}
\mathcal{I}[\tilde{U},\phi^{x}] 
= \int d\rho 
\left\{
(\dot{\tilde{U}})^{2}
+\tfrac{1}{3}g_{xy} \dot{\phi}^{x}  \dot{\phi}^{y}
-e^{2\tilde{U}}V_{\rm BB}
+(\omega/2)^{2}
\right\}\, .   
\end{equation}

\noindent
where the negative semidefinite black-brane potential, after an adequate
choice of normalization, is given by

\begin{equation}
-V_{\rm BB}(\phi,p) \ =\ a_{IJ}p^{I}p^{J}\, , 
\end{equation}

\noindent
where we denote by $p^{I}$ the electric charges of the string

\begin{equation}
p^{I} \sim \int_{S^{2}_{\infty}}a^{IJ}\star H_{J}\, .   
\end{equation}

\noindent
The Hamiltonian constraint (\ref{eq:hamiltonianconstraint}) becomes

\begin{equation}
\label{eq:Effe2a}
(\dot{\tilde{U}})^{2}
+\tfrac{1}{3}g_{xy}\dot{\phi}^{x} \dot{\phi}^{y}
+e^{2\tilde{U}}V_{\rm BB}
= 
(\omega/2)^{2}\, .
\end{equation}

If we define the {\em dual central charge} $\tilde{\cal Z}(\phi,p)$
by\footnote{This definition should be compared to that the standard central
  charge $\mathcal{Z}(\phi,q)=h^{I}q_{I}$.}

\begin{equation}
\tilde{\mathcal{Z}}(\phi,q)\equiv h_{I}p^{I}\, ,
\end{equation}

\noindent
it is possible to rewrite the black-brane potential in the form

\begin{equation}
-V_{\rm BB} \ =\  
\tilde{\mathcal{Z}}^{2}
+3g^{xy}\partial_{x}\tilde{\mathcal{Z}}\partial_{y}\tilde{\mathcal{Z}}\, ,
\end{equation}

\noindent
where Eq.~(\ref{eq:a-1}) has been used to obtain the last expression. Just as
it happens in the black-hole case, this form of the black-brane potential
allows us to rewrite the effective action Eq.~(\ref{eq:effectived5}) in a BPS
form, \textit{i.e.}~as a sum of squares up to a total derivative

\begin{equation}
\label{eq:effectived5BPS}
\mathcal{I}[\tilde{U},\phi^{x}] 
= \int d\rho 
\left\{
  \left(\dot{\tilde{U}}\pm e^{\tilde{U}}\tilde{\mathcal{Z}} \right)^{2}
  +\tfrac{1}{3}g_{xy} 
  \left( \dot{\phi}^{x} \pm 3 e^{\tilde{U}}\partial^{x}\tilde{\mathcal{Z}}\right)
  \left( \dot{\phi}^{y} \pm 3 e^{\tilde{U}}\partial^{y}\tilde{\mathcal{Z}}\right)
  \mp \frac{d~}{d\rho}\left(e^{\tilde{U}}\tilde{\mathcal{Z}}\right)
\right\}\, .   
\end{equation}

\noindent
The action is, then, extremized, and the second-order equations of motion that
follow from the action are satisfied when the first-order BPS equations

\begin{eqnarray}
\dot{\tilde{U}}
& = & 
\mp e^{\tilde{U}}\tilde{\mathcal{Z}}\, ,
\\
& & \nonumber \\
\dot{\phi}^{x} 
& = &
\mp 3 e^{\tilde{U}}\partial^{x}\tilde{\mathcal{Z}}\, .
\end{eqnarray}

Observe that the equations of motion that follow from the action do not set
the Hamiltonian to any particular value. Actually, these first-order equations
imply the Hamiltonian constraint for $\omega=0$, \textit{i.e.}~for extremal
strings.  It should be possible to show that the extremal strings that satisfy
the above equations are, precisely, the supersymmetric ones.

On the horizon of these solutions the dual central charge reaches a
stationary point

\begin{equation}
\label{eq:extremalcharge}
\left. \partial_{x}\tilde{\mathcal{Z}}\right|_{\rm \phi_{\rm h}}\ =\ 0\, .  
\end{equation}

\noindent
The above condition and the properties of real special geometry imply that the
black-brane potential also reaches a stationary point on the horizon. The
converse is not always true and we expect the existence of extremal,
non-supersymmetric black strings and we are going to construct some solutions
of this kind explicitly in the following sections.

The extremal supersymmetric strings of these theories saturate the
supersymmetric BPS bound, \textit{i.e.}

\begin{equation}
T_{p} =\tfrac{3}{4}| \tilde{\mathcal{Z}}(\phi_{\infty},p) |\, .  
\end{equation}

\noindent
On the horizon, the general relation between entropy density and black-brane
potential Eq.~(\ref{eq:extremalarea}) plus the particular property
Eq.~(\ref{eq:extremalcharge}) imply that the entropy density is determined by
the value of the dual central charge on the horizon (here $\tilde{p}=0$)

\begin{equation}
\label{eq:extremalareaN2d5}
\tilde{S}
= 
|\tilde{\cal Z}(\phi_{\rm h},p)|^{2}\, .
\end{equation}

There is a well-established procedure to construct all the extremal
supersymmetric strings of an ungauged $N=2,d=5$ supergravity coupled to $n$
vector multiplets \cite{Gauntlett:2002nw}: given $n+1$ spherically-symmetric
real harmonic functions on Euclidean $\mathbb{R}^{3}$

\begin{equation}
\label{eq:harmonicfunctions}
K^{I} = K^{I}_{\infty} +p^{I}\rho\, ,  
\end{equation}

\noindent
the fields of the supersymmetric solutions are implicitly given in terms of
these functions by the relations

\begin{equation}
\label{eq:stabil}
e^{-\tilde{U}}\ h^{I}(\phi ) \, =\, K^{I}\, .
\end{equation}

\noindent
We will denote the explicit expressions for the physical fields of the
solutions with the subscript $\mathrm{susy}$: $\tilde{U}_{\rm
  susy}=\tilde{U}_{\rm susy}(K)$, $\phi^{i}_{\rm susy}=\phi^{i}_{\rm
  susy}(K)$.


\subsection{A one-modulus model}
\label{sec:onemodulus}


In this section we are going to apply the formalism developed in the previous
sections to construct the black-string solutions of the simple model of
$N=2,d=5$ coupled to one vector multiplet whose black-hole solutions were
constructed in Ref.~\cite{Meessen:2011bd}. This model, which can be obtained
by dimensional reduction of minimal $d=6$ $N=(1,0)$ supergravity, is
determined by $C_{011}=1/3$. The hypersurface defined by
Eq.~(\ref{eq:hypersurface}) has to be covered by two coordinate patches that
determine two branches of the theory. We label these two branches by
$\sigma=\pm 1$. The relation between the projective coordinates $h$ and the
physical scalar $\phi$ in both branches is given by

\begin{equation}
  \begin{array}{rclrcl}
h_{(\sigma )}^{0} & = & e^{\sqrt{\frac{2}{3}}\phi}\, ,
\hspace{1cm}& 
h_{(\sigma )}^{1} & = &  \sigma\ e^{-\frac{1}{\sqrt{6}}\phi}\, , \\ 
& & & & & \\
h_{(\sigma )\, 0} & = & \tfrac{1}{3} e^{-\sqrt{\frac{2}{3}}\phi}\, ,
\hspace{1cm}& 
h_{(\sigma )\, 1} & = & \tfrac{2}{3} \sigma\ e^{\frac{1}{\sqrt{6}}\phi}\, . \\ 
\end{array}
\end{equation}

\noindent
The scalar metric $g_{\phi\phi}$ and the vector field strengths metric
$a_{IJ}$ take exactly the same values in both branches:

\begin{equation}
  \label{eq:11}
  g_{\phi\phi} \ =\ 1 \hspace{.5cm},\hspace{.5cm}
  a_{IJ} \ =\ \tfrac{1}{3}\left(
        \begin{array}{cc}
              e^{-2\sqrt{\frac{2}{3}}\phi} & 0 \\
              0 &  2 e^{\sqrt{\frac{2}{3}}\phi} \\
        \end{array}
        \right)\, ,
\end{equation}

\noindent
and, therefore, the bosonic parts of both models and their classical solutions
are identical. However, since the functions $h_{(\sigma )}^{I}(\phi)$ differ,
the fermionic structure and, therefore, the supersymmetry properties of a
given solution will be different in different branches. In particular, the
dual central charge is different in each branch:

\begin{equation}
\tilde{\mathcal{Z}}_{(\sigma )}  = \tfrac{1}{3}\left( p^{0}e^{-\sqrt{\frac{2}{3}}\phi}
+2 \sigma p^{1}e^{\frac{1}{\sqrt{6}}\phi}\right)\, .  
\end{equation}

The black-brane potential is identical in both branches because it is a
property of the bosonic part of the theory. It is given by

\begin{equation}
-V_{\rm BB}  = 
\tfrac{1}{3}\left[   
(p^{0})^{2}e^{-2\sqrt{\frac{2}{3}}\phi}
+2(p^{1})^{2}e^{\sqrt{\frac{2}{3}}\phi}
\right]\, ,
\end{equation}

\noindent
and it is extremized for

\begin{equation}
\phi_{\rm h} = 
\sqrt{\tfrac{2}{3}} \log{\left(\pm \sigma \frac{p^0}{p^1}\right)}\, ,
\end{equation}

\noindent
taking the value

\begin{equation}
\label{eq:bbpotentialonthehorizon}
-V_{\rm BB} (\phi_{\rm h},p) 
= 
\left[|p^{0}| (p^{1})^{2}\right]^{\frac{2}{3}}\, ,  
\end{equation}

\noindent
in all cases, while the dual central charge takes the value

\begin{equation}
\label{eq:dualcentralchargeonthehorizon}
\tilde{\cal Z}(\phi_{\rm h}, p)
= \tfrac{1}{3}(1\pm 2)\, \mathrm{sign}(p^{0})
\left[|p^{0}|(p^{1})^{2}\right]^{\frac{1}{3}}\, .  
\end{equation}

Since $\pm\sigma p^{0}/p^{1} >0$, the upper sign (which corresponds to the
supersymmetric case in the $\sigma$-branch, because it extremizes the dual
central charge) requires the following relation between the signs of the
charges $p^{I}$

\begin{equation}
\mathrm{sign}(p^{0}) = \sigma \mathrm{sign}(p^{1})\, ,
\end{equation}

\noindent
while the lower sign (which corresponds to non-supersymmetric extremal black
strings in the $\sigma$-branch) requires

\begin{equation}
\mathrm{sign}(p^{0}) = -\sigma \mathrm{sign}(p^{1})\, .
\end{equation}

We are going to construct the supersymmetric solutions of the $\sigma$-branch
next; the non-supersymmetric solutions of the $(-\sigma )$-branch will be
constructed at the same time.


\subsection{Supersymmetric and non-supersymmetric extremal solutions}
\label{sec:susysol}


The general prescription tells us that the extremal supersymmetric solutions
are given by two real harmonic functions of the form
Eq.~(\ref{eq:harmonicfunctions}), and are related to $\tilde{U}_{\rm susy}$
and $\phi_{\rm susy}$ by Eqs.~(\ref{eq:stabil}), which in this case take the
form

\begin{equation}
\label{eq:2}
K^{0} 
\ =\ 
e^{-\tilde{U}_{\rm susy}}\ e^{\sqrt{\frac{2}{3}}\, \phi_{\rm susy}}\, ,
\hspace{1cm}
K^{1} 
\ =\  
\sigma\ e^{-\tilde{U}_{\rm susy}}\ e^{\frac{-1}{\sqrt{6}}\phi_{\rm susy}} \; .
\end{equation}

\noindent
Then, $\tilde{U}_{\rm susy}$ and $\phi_{\rm susy}$ are given by

\begin{equation}
\label{eq:USUSY}
e^{-\tilde{U}_{\rm susy}} 
\ =\ 
\left[ K^{0}(K^{1})^{2} \right]^{1/3} \, ,
\hspace{1cm}
\phi_{\rm susy} 
\ =\  
\sqrt{\tfrac{2}{3}} \log\left(\sigma \frac{K^{0}}{K^{1}}\right)\; .
\end{equation}

\noindent
For these fields to be regular and well-defined, the harmonic functions
$K^{I}$ must satisfy several conditions\footnote{These restrictions can be
  read directly from Eqs.~(\ref{eq:2}).}:

\begin{enumerate}
\item[i)] They should not vanish at any finite, positive, value of $\rho$:
  this requirement relates the signs of the two constants that enter in each
  function $K^{I}$, $p^{I}$ and $K^{I}_{\infty}$:
\begin{equation}
\mathrm{sign}(K^{I}_{\infty})= \mathrm{sign}(p^{I})\, .  
\end{equation}
  
\item[ii)] For $\phi_{\rm susy}$ to be well-defined in the $\sigma$-branch

\begin{equation}
\mathrm{sign}(K^{0})=\sigma\ \mathrm{sign}( K^{1})\, ,
\end{equation}

\noindent
everywhere. This implies, in particular, that $\mathrm{sign}(p^{0})=\sigma
\mathrm{sign}(p^{1})$ which is the relation we found for the supersymmetric
critical points. Thus, there are two supersymmetric cases for each branch
which are disjoint in charge space: $\mathrm{sign}(p^{0})=+1,
\mathrm{sign}(p^{1})=\sigma$ and $\mathrm{sign}(p^{0})=-1,
\mathrm{sign}(p^{1})=-\sigma$.
\item[iii)] For $\tilde{U}_{\rm susy}$ to be well-defined ($e^{-\tilde{U}}>0$)
  we must have

\begin{equation}
K^{0}>0\, ,
\hspace{1cm}
\mathrm{sign}( K^{1}) = \sigma\, ,\,\,\, \Rightarrow\,\,\,
p^{0}>0\, ,\,\,\,
\mathrm{sign}(p^{1})\sigma>0\, .
\end{equation}

\end{enumerate}

It is, then, evident, that $K^{0}<0$ corresponds to the non-supersymmetric,
extremal case, which will be given by 

\begin{equation}
\label{eq:Unsusy}
e^{-\tilde{U}_{\rm nsusy}} 
\ =\ 
\left[ (-K^{0})(K^{1})^{2} \right]^{1/3}\, ,
\hspace{1cm}
\phi_{\rm nsusy} 
\ =\  
\sqrt{\tfrac{2}{3}} \log\left[\sigma \frac{(-K^{0})}{K^{1}}\right]\, .
\end{equation}

To summarize: the supersymmetric and the non-supersymmetric extremal solutions
can be written in this unified way:

\begin{equation}
\label{eq:Uex}
e^{-\tilde{U}_{\rm ext}} 
\ =\ 
\left[ |K^{0}|(K^{1})^{2} \right]^{1/3}\, ,
\hspace{1cm}
\phi_{\rm ext} 
\ =\  
\sqrt{\tfrac{2}{3}} \log\left|\frac{K^{0}}{K^{1}}\right|\, ,
\end{equation}

\noindent
with the harmonic functions given by

\begin{equation}
\label{eq:3}
K^{0}  
\ =\ 
\mathrm{sign}(p^{0})\, \left(e^{\sqrt{\frac{2}{3}}\phi_{\infty}}  +|p^{0}|\rho\right)\, ,
\hspace{1cm}
K^{1}  
\ =\ 
\sigma \left(
e^{-\frac{1}{\sqrt{6}}\phi_{\infty}}+ |p^{1}|\rho \right)\, .
\end{equation}

\noindent
The supersymmetric cases correspond to the signs $\mathrm{sign}(p^{0})> 0$
$\mathrm{sign}(p^{1})=\sigma$ and the non-supersymmetric ones to
$\mathrm{sign}(p^{0})<0$ and $\mathrm{sign}(p^{1})=-\sigma$.

The tension of these extremal solutions, defined in the $\rho\rightarrow 0$
limit by Eq.~(\ref{eq:generaltensionformula})
is given in all the cases by the manifestly positive quantity

\begin{equation}
\label{eq:tensionextremal}
T_{1} = \tfrac{1}{4}\left( |p^{0}|e^{-\sqrt{\frac{2}{3}}\phi}
+2 |p^{1}|e^{\frac{1}{\sqrt{6}}\phi}\right)\, ,  
\end{equation}

\noindent
which only equals the absolute value of the central charge when
$\mathrm{sign}(p^{0})= \sigma \mathrm{sign}(p^{1})$, which happens in the
supersymmetric cases.  Furthermore, in the supersymmetric cases, as we just
said, $\mathrm{sign}(p^{0})>0$ and

\begin{equation}
T_{1} = \tfrac{3}{4}\mathcal{Z}_{(\sigma )}(\phi_{\infty},p)\, .
\end{equation}

\noindent
In the non-supersymmetric cases, as one should expect, the mass is larger than
the central charges.

The entropy is given by the black-string potential on the horizon according to
the formula Eq.~(\ref{eq:extremalarea}). Then,
Eq.~(\ref{eq:bbpotentialonthehorizon}) tells us that the entropy density is,
in all extremal cases, given by 

\begin{equation}
\tilde{S} = \left[|p^{0}| (p^{1})^{2}\right]^{\frac{2}{3}}\, .  
\end{equation}

\noindent
Comparing with Eq.~(\ref{eq:dualcentralchargeonthehorizon}), we find that the
relation between the entropy density and the dual central charge on the
horizon Eq.~(\ref{eq:extremalareaN2d5}) only holds in the supersymmetric
cases. In the non-supersymmetric ones

\begin{equation}
\tilde{S} > |\tilde{\cal Z}(\phi_{\rm h}, p)|^{2} = \tfrac{1}{9} \tilde{S}\, . 
\end{equation}


\subsection{Non-extremal solutions}
\label{sec:nonextsol}


As in the black-hole case considered in Ref.~\cite{Meessen:2011bd}, the most
general solution can be obtained by direct integration using the fact that the
effective action is separable: defining the new variables

\begin{equation}
\label{eq:7}
x \ \equiv\ \tilde{U} -\sqrt{\tfrac{2}{3}}\, \phi \;\;\; ,\;\;\;
y \ \equiv\ \tilde{U} +\tfrac{1}{\sqrt{6}}\,\phi\, ,
\end{equation}

\noindent
the effective action Eq.~(\ref{eq:effectived5}) takes the form

\begin{equation}
\mathcal{I}[x,y] \; =\; 
\tfrac{1}{3}\int d\rho\
\left[\
(\dot{x})^{2}
\ +\ 2(\dot{y})^{2}
\ +\  (p^{0})^{2}e^{2x}
\ +\ 2 (p^{1})^{2}e^{2y}\
\right]\; ,   
\end{equation}

\noindent
and the equations of motion that follow from it can be integrated
immediately in full generality, giving

\begin{eqnarray}
\label{eq:Effe10a}
e^{-3\tilde{U}} & =& |p^{0}(p^{1})^{2}| \
\left(\frac{\sinh{(C\rho+D)}}{C}\right)^{2}
\left(\frac{\sinh{(A\rho+B)}}{A}\right)\, ,  \\
& &\nonumber \\
\phi & =& -\sqrt{\tfrac{2}{3}} 
\log{
\left\{
\left|\frac{p^{1}}{p^{0}}\right|
\left(\frac{A}{\sinh{(A\rho+B)}}\right)
\left(\frac{\sinh{(C\rho+D)}}{C}\right)
\right\}
}\, ,
\end{eqnarray}

\noindent
where $A,B,C$ and $D$ are (positive) integration constants. Their values are
related to the non-extremality parameter $\omega$ by the Hamiltonian
constraint Eq.~(\ref{eq:Effe2a})

\begin{equation}
\label{eq:Effe30}
2C^{2}\ +\ A^{2}\; =\;  3 (\omega/2)^{2}\, .  
\end{equation}

\noindent
The regularity of the solution imposes $A=C$. This constraint together with
the Hamiltonian constraint Eq.~(\ref{eq:Effe30}) implies 

\begin{equation}
A=C=\omega/2\, .
\end{equation}

\noindent
We are left with two constants, $B$ and $D$, that have to be expressed in
terms of the physical parameters of the solution by requiring $\tilde{U}(0)=0$
(asymptotic flatness) and $\phi(0)=\phi_{\infty}$:
which can be solved, yielding

\begin{eqnarray}
\label{eq:BD}
B 
& = & 
\log\left(\frac{\omega}{2|p^0|}e^{\sqrt{\frac{2}{3}}\phi_{\infty}}
+
\sqrt{1+\frac{\omega^2}{4|p^0|^2}e^{2\sqrt{\frac{2}{3}}\phi_{\infty}}}\right)\, ,
\\
& & \nonumber \\
D 
& = & 
\log\left(\frac{\omega}{2|p^1|}e^{-\frac{1}{\sqrt{6}}\phi_{\infty}}
+
\sqrt{1+\frac{\omega^{2}}{4|p^1|^2}e^{-\frac{2}{\sqrt{6}}\phi_{\infty}}}\right)\, .
\end{eqnarray}

\noindent
The tension is given by

\begin{equation}
\label{eq:mass}
T_{p}
= 
-\tfrac{1}{8}\omega +\tfrac{1}{8}\sqrt{\omega^{2}+4\left( p^{0}\right)^{2}
  e^{-2\sqrt{\frac{2}{3}}\phi_{\infty}}}
+\tfrac{1}{4}\sqrt{\omega^{2}+4\left(p^{1}\right)^{2} e^{\sqrt{\frac{2}{3}}\phi_{\infty}}}\, .
\end{equation}

\noindent
When the charges vanish we recover the Schwarzschild branes' tension
$T_{p}=|\omega|/2$. Taking $\omega=0$ we obtain the tension of all the
extremal cases Eq.~(\ref{eq:tensionextremal}).  This equation can be inverted
in order to explicitly identify the different extremal limits and the
correspondent mass, but the expression is very involved, so we will analyze
the extremal limits from Eq.~(\ref{eq:mass}).

The entropy density is given by 

\begin{eqnarray}
\label{eq:S}
\tilde{S} = \left| p^0 (p^1)^2\right|^{\frac{2}{3}}
\left(
\frac{\omega}{2|p^1|}e^{-\frac{1}{\sqrt{6}}\phi_{\infty}}
+\sqrt{1+\frac{\omega^2}{4|p^1|^2}e^{-\frac{2}{\sqrt{6}}\phi_{\infty}}}
\right)^{\frac{4}{3}} 
\left(
\frac{\omega}{2|p^0|}e^{\sqrt{\frac{2}{3}}\phi_{\infty}} 
+\sqrt{1+\frac{\omega^2}{4|p^0|^2}e^{2\sqrt{\frac{2}{3}}\phi_{\infty}}}
\right)^{\frac{2}{3}}\, .
\end{eqnarray}

\noindent
Taking the extremal limit $\omega\rightarrow 0$, we recover the expression
already found for the extremal case. The Hawking temperature can be found
using the relation between the entropy density, the temperature and the
non-extremality parameter Eq.~(\ref{eq:relation}).

\section*{Acknowledgments}

The authorswould like to thank R.~Emparan and P.~Meessen for useful
conversations. This work has been supported in part by the Spanish Ministry of
Science and Education grant FPA2009-07692, the Comunidad de Madrid grant
HEPHACOS S2009ESP-1473 and the Spanish Consolider-Ingenio 2010 program CPAN
CSD2007-00042. The work of CSS has been supported by a JAE-predoc grant JAEPre
2010 00613.  TO wishes to thank M.M.~Fern\'andez for her permanent support.

\appendix

\section{Black branes versus black holes: dimensional reduction}
\label{sec-dimred}

It is sometimes useful to consider the toroidal compactification of flat
$p$-branes over the $p$ spatial worldvolume directions to get a
$(d-p)=(\tilde{p}+4)$-black-hole solution. This is how the first $p$-brane
solutions were constructed \cite{Horowitz:1991cd}.

Let us consider the $d$-dimensional action Eq.~(\ref{eq:daction}) and the
ansatz 

\begin{equation}
\begin{array}{rcl}
ds^{2}_{(d)} 
& = &
K^{-\frac{2}{\tilde{p}+2}}ds^{2}_{(\tilde{p}+4)}
-K^{\frac{2}{p}}d\vec{y}^{\, 2}_{(p)}\, ,
\\
& & \\     
ds^{2}_{(\tilde{p}+4)}
& = & 
g_{\mu\nu}dx^{\mu}dx^{\nu}\, ,
\\
& & \\     
A^{\Lambda}_{(p+1)\, \mu\ y_{1}\cdots y_{p}}
& = & 
A^{\Lambda}{}_{\mu}\, ,
\end{array}
\end{equation}

\noindent
where the $(d-p)=(\tilde{p}+4)$-dimensional metric $g_{\mu\nu}$, 1-forms
$A^{\Lambda}{}_{\mu}$, worldvolume element $K$ and scalars $\phi^{i}$ are all
independent of the worldvolume coordinates $\vec{y}_{(p)}$.

The dimensionally-reduced theory is governed by the action

\begin{equation}
\label{eq:d-paction}
\mathcal{I}[g,A^{\Lambda},\phi^{i},K]
=
\int d^{\tilde{p}+4}x \sqrt{|g|}
\left\{
  R +\tfrac{(d-2)}{p(\tilde{p}+2)} (\partial\log{K})^{2}
+\mathcal{G}_{ij} \partial_{\mu} \phi^{i} \partial^{\mu} \phi^{j} 
+2 K^{-2\frac{(\tilde{p}+1)}{(\tilde{p}+2)}}
I_{\Lambda \Sigma} F^{\Lambda} \cdot F^{\Sigma}
\right\}\, ,
\end{equation}

\noindent
where $F^{\Lambda}=dA^{\Lambda}$ are 2-form field strengths.

To search for the static, spherically-symmetric black-hole solutions of this
model, we can use the $(d-p)=(\tilde{p}+4)$-dimensional version of the FGK
formalism given in Ref.~\cite{Meessen:2011bd} and assume that the black-hole
metric will be given by 

\begin{equation}
\label{eq:conformastatic}
 ds_{(\tilde{p}+4)}^{2} 
= 
e^{2U_{(\tilde{p}+4)}}dt^{2} 
-e^{-\frac{2}{\tilde{p}+1}U_{(\tilde{p}+4)}}
\gamma_{(\tilde{p}+3)\, \underline{m}\,\underline{n}}
dx^{\underline{m}} dx^{\underline{m}}\, , 
\end{equation}

\noindent
where $\gamma_{(\tilde{p}+3)\, \underline{m}\,\underline{n}}$ is the
background transverse metric given in Eq.~(\ref{eq:backgroundtransversemetric}). The
effective action controlling the dynamics of the black-hole warp factor
$U_{(\tilde{p}+4)}$, the worldvolume element $K$ and the scalars $\phi^{i}$ is
\cite{Meessen:2011bd}

\begin{equation}
\label{eq:Effe40}
\mathcal{I}[U_{(\tilde{p}+4)},\phi^{i},K] 
= \int d\rho 
\left\{
(\dot{U}_{(\tilde{p}+4)})^{2}
+\tfrac{(\tilde{p}+1)}{(\tilde{p}+2)}
\left[
\tfrac{(d-2)}{p(\tilde{p}+2)} K^{-2}(\dot{K})^{2}
+\mathcal{G}_{ij}\dot{\phi}^{i}\dot{\phi}^{j} \right]
-e^{2U_{(\tilde{p}+4)}}V_{\rm bh}
\right\}\, ,   
\end{equation}

\noindent
where the black-hole potential is given, up to the normalization constant
$\alpha$, by

\begin{equation}
V_{\rm bh} \, =\, 2\alpha^{2}\ \tfrac{(\tilde{p}+1)}{(\tilde{p}+2)}\ 
K^{2\frac{(\tilde{p}+1)}{(\tilde{p}+2)}}I^{\Lambda\Sigma}q_{\Lambda}q_{\Sigma}\, .
\end{equation}

\noindent
The Hamiltonian constraint takes the form

\begin{equation}
\label{eq:HamBsquare}
(\dot{U}_{(\tilde{p}+4)})^{2}
+\tfrac{(\tilde{p}+1)}{(\tilde{p}+2)}
\left[
\tfrac{(d-2)}{p(\tilde{p}+2)} K^{-2}(\dot{K})^{2}
+\mathcal{G}_{ij}\dot{\phi}^{i}\dot{\phi}^{j} \right]
+e^{2U_{(\tilde{p}+4)}}V_{\rm bh}= (\omega/2)^{2}\, ,
\end{equation}

\noindent
where $\omega$ is the non-extremality parameter in the background transverse
metric ($\omega=2\mathcal{B}$ in Ref.~\cite{Meessen:2011bd}).

The equations of motion for $U_{(\tilde{p}+4)}$ and $K$ are, respectively

\begin{eqnarray}
\ddot{U}_{(\tilde{p}+4)}  
+e^{U_{(\tilde{p}+4)}}V_{\rm bh}
& = & 
0\, ,
\\
& & \nonumber \\
\tfrac{(d-2)}{p(\tilde{p}+2)} \frac{d^{2}~}{d\rho^{2}}\log{K}
+e^{U_{(\tilde{p}+4)}}V_{\rm bh}
& = & 
0\, ,
\end{eqnarray}

\noindent
and their difference can be solved for $K$ as a function of
$U_{(\tilde{p}+4)}$ and two integration constants $a$ and $b$, giving

\begin{equation}
K= e^{\frac{p(\tilde{p}+2)}{(d-2)} U_{(\tilde{p}+4)} +a\rho+b}\, .  
\end{equation}

\noindent
For simplicity we normalize $K$ at spatial infinity to $1$ by setting $b=0$
and for latter convenience we redefine the integration constant $a=
-\frac{p(\tilde{p}+2)}{2(d-2)}\gamma$, so

\begin{equation}
K= e^{\frac{p(\tilde{p}+2)}{(d-2)}
  (U_{(\tilde{p}+4)}-\frac{1}{2}\gamma\rho)}\, .
\end{equation}

Using this result to eliminate $K$ from the equations of motion, we arrive at
a set of equations of motion that  can be derived from the effective action 
Eq.~(\ref{eq:geodesicaction}) upon the identifications

\begin{eqnarray}
\tilde{U} 
& \equiv &
\tfrac{(p+1)(\tilde{p}+2)}{(d-2)}U_{(\tilde{p}+4)}   
-\tfrac{p(\tilde{p}+1)}{2(d-2)}\gamma\rho\, ,
\\
& & \nonumber \\
V_{\rm BB}
& \equiv &
\tfrac{(p+1)(\tilde{p}+2)}{(d-2)}
V_{\rm bh}\, .
\end{eqnarray}


\section{Some known families of black-brane solutions}
\label{sec-known}

In this appendix we review several well-known families of black-brane
solutions in order to gain intuition and understand better the general setup
proposed in this paper.


\subsection{Schwarzschild black $p$-branes}

These solutions are obtained by trivial oxidation of the
$(\tilde{p}+4)$-dimensional generalization of the Schwarzschild solution
\cite{Tangherlini:1963bw}

\begin{equation}
\label{eq:dSchwar}
ds_{(\tilde{p}+4)}^{2}  = 
W\, dt^{2}-W^{-1} dr^{2}-r^{2}d\Omega_{(\tilde{p}+2)}^{2}\, , 
\hspace{1cm}
W=1+{\displaystyle\frac{\omega}{r^{\tilde{p}+1}}}\, ,
\end{equation}

\noindent
where $d\Omega_{(\tilde{p} +2)}^{2}$ is the metric of the
$(\tilde{p}+2)$-sphere of unit radius.  The oxidation to $d=p+\tilde{p}+4$
dimensions gives the direct product of the above metric with the
$p$-dimensional Euclidean metric $d\vec{y}_{(p)}^{\, 2}$:

\begin{equation}
\label{eq:dSchwarbb}
ds_{(d)}^{2}  = 
W\, dt^{2}-d\vec{y}_{(p)}^{\, 2}-W^{-1} dr^{2}-r^{2}d\Omega_{(\tilde{p}+2)}^{2}\, .
\hspace{1cm}
W=1+{\displaystyle\frac{\omega}{r^{\tilde{p}+1}}}\, .
\end{equation}

These metrics, which are asymptotically (i.e.~at $r\rightarrow +\infty$) flat
in the directions orthogonal to the brane's worldvolume, have an event horizon
at $r^{\tilde{p}+1}=-\omega$ (we take $\omega<0$) that hides any possible
curvature singularity at lower values of $r$. The first coefficient in the
expansion of $g_{tt}$ ($W$) in $1/r^{\tilde{p}+1}$ is the mass of the black
hole in $(d-p)$ dimensions\footnote{We choose the mass units so as to get a
  convenient coefficient.}:

\begin{equation}
W \sim 1 -\frac{2M}{r^{\tilde{p}+1}}\, ,  
\end{equation}
 
\noindent
and can be taken as the definition of the energy per unit of worldvolume
(tension) of the black $p$-brane in $d$ dimensions

\begin{equation}
T_{p}=M=-\omega/2\, .  
\end{equation}

A more general definition can be given, following
Ref.~\cite{Myers:1999psa}\footnote{We would like to thank R.~Emparan for his
  clarification on this point.}: if we expand the spacetime metric in the weak
field limit into the asymptotic metric (Minkowski $\eta_{\mu\nu}$) and a
perturbation $h_{\mu\nu}$
 
\begin{equation}
\label{eq:ctensor1}
g_{\mu\nu}=\eta_{\mu\nu}+h_{\mu\nu}\, ,
\hspace{1cm}
h_{\mu\nu}=  \frac{c_{\mu\nu}}{r^{\tilde{p}+1}}\, ,  
\end{equation}

\noindent
where $c_{\mu\nu}$ is a constant tensor, then, the $p$-brane's stress-enery
tensor $t_{ab}$ ($a,b=0,\cdots,p$) is given by

\begin{equation}
\label{eq:stressenergytensor1} 
t_{ab}= -\frac{\omega_{\tilde{p}+2}}{16\pi G_{N(d)}} 
\left[(\tilde{p}+1)c_{ab} +\eta_{ab}\eta^{cd}c_{cd}\right]\, , 
\end{equation}

\noindent
where $\omega_{\tilde{p}+2}$ is the volume of the unit $(\tilde{p}+2)$-sphere
and $G_{N(d)}$ is the $d$-dimensional Newton constant. The component $t_{00}$
gives the tension $T_{p}$ and we recover the above value choosing units such
that

\begin{equation}
\label{eq:units}
\frac{\omega_{\tilde{p}+2}(\tilde{p}+2)}{8\pi G_{N(d)}}=1\, .  
\end{equation}

The definition of the constant tensor $c_{\mu\nu}$ will change slightly when
we change coordinates, but the expression Eq.~(\ref{eq:stressenergytensor1})
will still be valid. The tension will coincide with the mass of the black hole
that one obtains by dimensional reduction if one uses carefully the relation
between the $d$ and the $(\tilde{p}+4)$-dimensional Newton constant.

The angular part of the metric remains finite in the limit $r\rightarrow
(-\omega)^{\frac{1}{\tilde{p}+1}}$ and the volume of the
$(\tilde{p}+2)$-spheres converges to a finite value there:
$\omega^{\tilde{p}+2}$ times the volume of the unit $(\tilde{p}+2)$-sphere.
Redefining the radial coordinate to one, $R$, which
vanishes on the horizon

\begin{equation}
r^{\tilde{p}+1} = \left(\frac{\tilde{p}+1}{2} \right)^{2}
(-\omega)^{\frac{\tilde{p}-1}{\tilde{p}+1}} R^{2} -\omega\, ,  
\end{equation}

\noindent
the metric, in the near-horizon limit takes the form

\begin{equation}
ds_{(d)}^{2} \sim 
\left(\frac{\tilde{p}+1}{2} \right)^{2}
(-\omega)^{-\frac{2}{\tilde{p}+1}} 
R^{2}dt^{2} -dR^{2}
-d\vec{y}^{\, 2}_{(p)}
-(-\omega)^{\frac{2}{\tilde{p}+1}}d\Omega^{2}_{(\tilde{p}+2)}\, ,
\end{equation}

\noindent
which is the direct product of a Rindler space (in the time-radial
directions), a $p$-dimensional Euclidean space (the brane's worldvolume) and a
$(\tilde{p}+2)$-sphere of radius $(-\omega)^{\frac{1}{\tilde{p}+1}}$. Now
Wick-rotating the time coordinate and requiring the time-radial part of the
metric metric to be free of conical singularities, we find that the Euclidean
time must be compact with period (inverse Hawking temperature)

\begin{equation}
\beta = \frac{ 4\pi (-\omega)^{\frac{1}{\tilde{p}+1}}}{\tilde{p}+1}\, .
\end{equation}

\noindent
The volume of the $(\tilde{p}+2)$-dimensional sections of constant $t$ and
$\vec{y}^{\, 2}_{(p)}$ of the horizon is given by

\begin{equation}
\frac{A_{{\rm h}\, (\tilde{p}+2)}}{\omega_{(\tilde{p}+2)}} = 
(-\omega)^{\frac{\tilde{p}+2}{\tilde{p}+1}}\, ,
\end{equation}

\noindent
where $\omega_{(\tilde{p}+2)}$ is the volume of the unit
$(\tilde{p}+2)$-sphere. If the $p$-dimensional spacelike worldvolume were
compact, then the above quantity would be equal to the quotient of the
$(d-1)$-dimensional constant-time sections of the horizon and the
$p$-dimensional spacelike worldvolume, and, therefore, up to numerical
constants (in our conventions), it can be interpreted as the entropy density
by unit of worldvolume. We will denote this quantity by $\tilde{S}$ and, thus,

\begin{equation}
\label{eq:entropydensity}
\tilde{S} \equiv   
\frac{A_{{\rm h}\, (\tilde{p}+2)}}{\omega_{(\tilde{p}+2)}} = 
(-\omega)^{\frac{\tilde{p}+2}{\tilde{p}+1}}\, .
\end{equation}

In this work we use a radial coordinate $\rho$ for which the event horizon
lyes at $+\infty$ and spatial infinity at $\rho\rightarrow 0$ and which is
related to $r$ by two consecutive changes of coordinates: first $r\rightarrow
z$

\begin{equation}
r 
= 
z \left(1-\frac{\omega/4}{z^{\tilde{p}+1}}\right)^{\frac{2}{\tilde{p}+1}}\, , 
\end{equation}

\noindent
which brings the metric into an isotropic (in transverse space) form.  For the
above Schwarzschild black $p$-branes, this isotropic form of the metric is

\begin{equation}
ds^{2}
=
\frac{W_{+}^{2}}{W_{-}^{2}}dt^{2}
-d\vec{y}_{(p)}^{\, 2}  
-W_{-}^{\frac{4}{\tilde{p}+1}}
\left[dz^{2}+z^{2} d\Omega_{(\tilde{p}+1)}^{2} \right]\, ,
\hspace{1cm}
W_{\pm} = 1 \pm \frac{\omega/4}{z^{\tilde{p}+1}}\, .
\end{equation}

\noindent
The second coordinate change $z\rightarrow \rho$ is given by

\begin{equation}
z 
= 
\left(\frac{\omega/4}{\tanh{\frac{\omega}{4}\rho}}
\right)^{\frac{1}{\tilde{p}+1}}\, ,
\end{equation}

\noindent
and brings the metric into the final form

\begin{equation}
ds^{2}
=
e^{\omega\rho}dt^{2}
-d\vec{y}_{(p)}^{\, 2}  
-e^{-\frac{1}{\tilde{p}+1}\omega\rho}
\gamma_{(\tilde{p}+3)\, \underline{m}\underline{n}}  dx^{m} dx^{n}\, ,
\end{equation}

\noindent
where the background transverse metric is

\begin{equation}
\gamma_{(\tilde{p}+3)\, \underline{m}\underline{n}}  dx^{m} dx^{n}
= 
\left(\frac{ \omega/2}{\sinh{\left(\frac{\omega}{2} \rho\right)}} \right)^{\frac{2}{\tilde{p}+1}}
\left[ 
\left( \frac{\omega/2}{\sinh{\left(\frac{\omega}{2} \rho \right)}}\right)^2
\frac{d\rho^2}{(\tilde{p}+1)^2}
+ d\Omega^{2}_{(\tilde{p}+2)} 
\right]\, .	 
\end{equation}

This background transverse metric is the $p$-brane generalization of the
$d$-dimensional generalization given in Ref.~\cite{Meessen:2011bd} of the
4-dimensional black-holes background transverse metric given in
Ref.~\cite{Ferrara:1997tw}).

At spatial infinity $\rho \rightarrow 0$, the exponentials that appear in the
metric go to $1$ (because $\omega<0$) and the background transverse metric
approaches

\begin{equation}
\label{eq:asympgamma}
\gamma_{(\tilde{p}+3)\, \underline{m}\underline{n}}  dx^{m} dx^{n}
\sim
\rho^{-\frac{2}{\tilde{p}+1}} \left[ \rho^{-2}\frac{d\rho^{2}}{(\tilde{p}+1)^{2}} 
+d\Omega^{2}_{(\tilde{p}+2)}  \right]\, ,
\end{equation}

\noindent
which is nothing but the $(\tilde{p}+3)$-dimensional Euclidean metric as can
be seen with the coordinate change $\rho^{-\frac{1}{\tilde{p}+1}}
=\varrho$. 

In these coordinates the tension is computed using
Eq.~(\ref{eq:stressenergytensor1}) where the constant tensor $c_{\mu\nu}$ is
now defined by

\begin{equation}
\label{eq:ctensor2}
h_{\mu\nu} = c_{\mu\nu}\rho\, .  
\end{equation}

In the near-horizon limit, the angular part of the background transverse
metric behaves as

\begin{equation}
\label{eq:nearhorizonangular}
\sim e^{\frac{1}{\tilde{p}+1}\omega \rho}
(-\omega)^{\frac{2}{\tilde{p}+1}}d\Omega^{2}_{(\tilde{p}+2)}\, ,
\end{equation}

\noindent
and becomes singular (shrinks to zero volume) on the horizon. This behavior is
compensated by the divergence of the factor $e^{-\frac{1}{\tilde{p}+1}\omega
  \rho}$ which sits in front of it, so that the result
Eq.~(\ref{eq:entropydensity}) is recovered.

In the same limit, the time-radial part of the metric behaves, after a
rescaling of the radial coordinate, as

\begin{equation}
\exp{\left(-\frac{\tilde{p}+1}{(-\omega)^{\frac{1}{\tilde{p}+1}}}\varrho\right)}
\left[dt^{2} -d\varrho^{2} \right]
=
e^{- \frac{4\pi}{\beta} \varrho}
\left[dt^{2} -d\varrho^{2} \right]
\, ,
\end{equation}

\noindent
from which one can easily read the temperature.

The tension, temperature and entropy density of Schwarzschild black $p$-branes
are the same as the mass, temperature and entropy of the Schwarzschild black
hole related to them by toroidal compactification. For more complex solutions,
the tension of the $p$-brane and the mass of the corresponding black hole will
be different, but the temperature and entropy will have the same values.


\subsection{RN black $p$-branes}

Our next example will be that of ``Reissner-Nordstr\"om'' $p$-branes, which
are charged solutions of the following action, which does not include scalar
fields:

\begin{equation}
\mathcal{I}[g_{\mu\nu},A_{(p+1)\, \mu_{1}\cdots \mu_{p+1}}]
=
\int d^{d}x \sqrt{|g|}
\left\{
R  +\frac{(-1)^{p+1}}{2\cdot(p+2)!} F_{(p+2)}^{2}
\right\}\, .
\end{equation}



Solutions describing static, flat, black $p$-branes charged with respect to
the $(p+1)$-form potential $A_{(p+1)}$, lying in the directions parametrized
by $\vec{y}_{(p)}\equiv (y_{1},\cdots, y_{p})$ were constructed in
Ref.~\cite{Horowitz:1991cd} and they are given by

\begin{equation} 
\label{eq:HoS}
\begin{array}{rcl}
ds_{(d)}^{2} 
& = & 
H^{-\frac{2}{p+1}}\left[Wdt^{2} -d\vec{y}^{\, 2}_{(p)} \right] 
-H^{\frac{2}{\tilde{p}+1}}
\left[W^{-1}dr^{2} +r^{2}d\Omega_{(\tilde{p} +2)}^{2}\right]\, , \\
& & \\
A_{(p+1)\,  t\underline{y}^{1}\cdots \underline{y}^{p}} 
& = &  
\alpha \left(H^{-1}-1\right)\, ,
\hspace{1cm}
H=1+ {\displaystyle\frac{h}{r^{\tilde{p}+1}}}\, ,
\hspace{1cm}
W=1+{\displaystyle\frac{\omega}{r^{\tilde{p}+1}}}\, ,
\end{array}
\end{equation}

\noindent
where the integration constants $\omega, h$ and $\alpha$ are related by

\begin{equation}
\label{eq:cdef}
\alpha^{2} = 2c\left(1-\omega/h \right)\, ,
\hspace{1cm}
c\equiv \frac{d-2}{(p+1)(\tilde{p}+1)}\, .  
\end{equation}

We are going to assume that $\omega \leq 0$ and $h\geq 0$, but otherwise
arbitrary. This is consistent with $\alpha^{2}\geq 0$ for all the possible
values of $\omega$ and $h$.

These solutions generalize the $d$-dimensional Reissner-Nordstr\"om black-hole
solutions \cite{Myers:1986un} which are the $p=0$ case. In all cases $\omega$
(which has to be non-positive for the solutions to have a regular event
horizon) plays the role of non-extremality parameter: when $\omega=0$ ($W=1$)
the solutions become extremal and, in some cases, supersymmetric. In this
limit $H$ can be replaced by an arbitrary harmonic function in the
$(\tilde{p}+3)$-dimensional transverse space, although only some choices give
physically meaningful solutions. When $h=0$ the solutions describe the
Schwarzschild black branes discussed in the previous section.

As they stand, these solutions are asymptotically flat in the directions
orthogonal to the worldvolume and have an event horizon at
$r^{\tilde{p}+1}=-\omega$ that hides any possible curvature singularity at
lower values of $r$. 

The tension can be computed using Eqs.~(\ref{eq:ctensor2}) and
(\ref{eq:stressenergytensor1}) and is given, in the units of
Eq.~(\ref{eq:units}) by

\begin{equation}
T_{p}= \frac{(d-2)}{(p+1)(\tilde{p}+2)}h -\omega/2\, ,
\end{equation}

\noindent 
and, again, it coincides with the mass of the
$(d-p)=(\tilde{p}+4)$-dimensional black hole that one gets by toroidal
compactification of the black brane over the $p$ worldvolume
directions. Observe that our choice of signs for $\omega$ and $h$ guarantees
the positivity of $T_{p}$.  Observe that the worldvolume element, given
locally by

\begin{equation}
K = H^{-\frac{p}{p+1}}\, ,  
\end{equation}

\noindent 
which becomes a scalar in $(d-p)$ dimensions, is normalized to $1$
at infinity.

To study the near-horizon limit we first redefine the radial coordinate

\begin{equation}
r^{\tilde{p}+1} = 
\left(\frac{\tilde{p}+1}{2} \right)^{2}
(-\omega)^{\frac{\tilde{p}-1}{\tilde{p}+1}} 
\left(1-\frac{h}{\omega}\right)^{-\frac{2}{\tilde{p}+1}} R^{2} -\omega\, ,  
\end{equation}

\noindent
after which the metric, in that limit, takes the form

\begin{equation}
ds_{(d)}^{2} 
\sim 
\left(\frac{\tilde{p}+1}{2} \right)^{2}
(-\omega)^{-\frac{2}{\tilde{p}+1}} 
\left(1-\frac{h}{\omega}\right)^{-2c} R^{2}dt^{2} -dR^{2}
-\left(1-\frac{h}{\omega}\right)^{-\frac{2}{p+1}}d\vec{y}^{\, 2}_{(p)}
-(h-\omega)^{\frac{2}{\tilde{p}+1}}d\Omega^{2}_{(\tilde{p}+2)}\, ,
\end{equation}

\noindent
which is the direct product of a Rindler space (in the time-radial
directions), a $p$-dimensional Euclidean space (the brane's worldvolume) and a
$(\tilde{p}+2)$-sphere of radius $(h-\omega)^{\frac{1}{\tilde{p}+1}}$. By the
usual argument, we find that the inverse Hawking temperature $\beta$ and the
entropy density $\tilde{S}$ are given by

\begin{equation}
\beta = \frac{ 4\pi (-\omega)^{\frac{1}{\tilde{p}+1}}}{\tilde{p}+1}
\left(1-\frac{h}{\omega}\right)^{c}\, ,
\hspace{1cm}
\tilde{S} = 
(h-\omega)^{\frac{\tilde{p}+2}{\tilde{p}+1}}\, ,
\end{equation}

If we change the radial coordinate from $r$ to $\rho$, defined in the previous
section, we find that the solution is now given by 

\begin{equation} 
\label{eq:HoS-2}
\begin{array}{rcl}
ds_{(d)}^{2} 
& = & 
\hat{H}^{-\frac{2}{p+1}}\left[e^{\frac{p}{p+1}\omega\rho}dt^{2} 
-e^{-\frac{1}{p+1}\omega\rho}d\vec{y}^{\, 2}_{(p)} \right] 
-\hat{H}^{\frac{2}{\tilde{p}+1}}
\gamma_{(\tilde{p}+3)\, \underline{m}\underline{n}}  dx^{m} dx^{n}\, ,\\
& & \\
A_{(p+1)\,  t\, y^{1}\cdots y^{p}} 
& = &  
\alpha \left(e^{-\frac{1}{2}\omega\rho }\hat{H}^{-1}-1\right)\, ,
\hspace{1cm}
\hat{H}=\cosh{\left(\frac{\omega}{2}\rho\right)}
+\left({\displaystyle\frac{2h}{\omega}}
-1 \right)\sinh{\left(\frac{\omega}{2}\rho\right)}\, ,
\end{array}
\end{equation}

\noindent
where the integration constants satisfy the same relations as before and where
the background transverse metric $\gamma_{(\tilde{p}+3)\,
  \underline{m}\underline{n}}$ is defined in
Eq.~(\ref{eq:backgroundtransversemetric}).


\subsubsection{FGK coordinates}

Based on the form of this metric, we can make the following ansatz for the
metrics of all charged black $p$-branes

\begin{equation}
ds_{(d)}^{2}
= 
e^{\frac{2}{p+1}\tilde{U}}
\left[
W^{\frac{p}{p+1}} dt^{2}
-W^{-\frac{1}{p+1}}d\vec{y}^{\, 2}_{(p)}
\right]  
-e^{-\frac{2}{\tilde{p}+1}\tilde{U}}
\gamma_{(\tilde{p}+3)\, \underline{m}\underline{n}}  dx^{m} dx^{n}\, .
\end{equation}

\noindent
For RN black $p$-branes

\begin{equation}
e^{-\tilde{U}} = \hat{H}\, ,
\hspace{1cm}
W= e^{\omega \rho}\, ,  
\end{equation}

\noindent
and for Schwarzschild black $p$-branes

\begin{equation}
e^{-\tilde{U}} = e^{-\frac{\omega}{2}\rho}\, ,
\hspace{1cm}
W= e^{\omega \rho}\, .
\end{equation}

In general, in the near-horizon limit, the angular part of the transverse
metric behaves as in Eq.~(\ref{eq:nearhorizonangular}), which means in that
black $p$-branes with regular horizons $\tilde{U}$ behaves as

\begin{equation}
\label{eq:asympU}
\tilde{U} \sim C+\frac{\omega}{2}\rho\, ,  
\end{equation}

\noindent
and, therefore, we get

\begin{equation}
\tilde{S}
= \left(- e^{-C}\omega\right)^{\frac{\tilde{p}+2}{\tilde{p}+1}}\, .
\end{equation}

\noindent
We can invert this relation to identify $C$ in terms of physical constants

\begin{equation}
e^{C} = -\omega \tilde{S}^{-\frac{\tilde{p}+1}{\tilde{p}+2}}\, . 
\end{equation}

\noindent
Taking into account this fact, in order for the worldvolume metric to be
regular in this limit, $\tilde{U}$ and $W$ must behave as\footnote{As shown in
  Section~\ref{sec:branesformalism}, in all cases $W=e^{\gamma\rho}$ for
  certain constant $\gamma$. As we see here, regularity of the horizon
  requires $\gamma=\omega$.}

\begin{equation}
\label{eq:asympW}
e^{\tilde{U}}
\sim
(-\omega) \tilde{S}^{-\frac{\tilde{p}+1}{\tilde{p}+2}} e^{\frac{\omega}{2}\rho}\, ,
\hspace{1cm}
W \sim e^{\omega\rho}\, ,  
\end{equation}

\noindent
where have chosen arbitrarily a normalization constant. The general metric for
regular $p$-branes is, therefore, given by Eq.~(\ref{eq:generalmetric}).

Combining these facts we find that the near-horizon limit of the time-radial
part of the metric can be brought into the Rindler-like form

\begin{equation}
\sim e^{\frac{2}{p+1}C} 
\exp{\left(-\frac{(\tilde{p}+1)
      e^{Cc}}{(-\omega)^{\frac{1}{\tilde{p}+1}}}\varrho \right)}  
\left[dt^{2}-d\varrho^{2}\right]
=e^{-\frac{4\pi}{\beta}\rho }\left[dt^{2}-d\varrho^{2}\right]\, ,
\end{equation}

\noindent
where $c$ is the constant defined in Eq.~(\ref{eq:cdef}), from which we find

\begin{equation}
\beta = \frac{4\pi (-\omega)^{\frac{1}{\tilde{p}+1}}}{(\tilde{p}+1)e^{Cc}} \, .
\end{equation}

The non-extremality parameter is related to the temperature and entropy by

\begin{equation}
\label{eq:relation}
(-\omega)^{\frac{1}{p+1}} = \tfrac{4\pi}{\tilde{p}+1} T 
\tilde{S}^{\frac{(d-2)}{(p+1)(\tilde{p}+2)}}\, .
\end{equation}


\subsubsection{Extremal limit}
\label{sec-extremal}

In the extremal limit $W=1$ and the transverse background metric takes the
form in Eq.~(\ref{eq:asympgamma}), which is just the
$(\tilde{p}+3)$-dimensional Euclidean metric as can be seen with the
coordinate change $\rho^{-\frac{1}{\tilde{p}+1}} =\varrho$. Then, in the
near-horizon limit, for the horizon to be regular, $\tilde{U}$ must approach

\begin{equation}
\label{eq:Uasymp}
e^{\tilde{U}}
\sim
\tilde{S}^{-\frac{\tilde{p}+1}{\tilde{p}+2}} \rho^{-1}\, .
\end{equation}

Finally, in these coordinates, the tension is given by 

\begin{equation}
\label{eq:generaltensionformula}
T_{p} = -\frac{1}{(p+1)(\tilde{p}+2)}
\left[(d-2)\tilde{u} +p(\tilde{p}+1)\omega/2\right]\, ,  
\end{equation}

\noindent
where $\tilde{u}$ is defined in the $\rho\rightarrow 0$ limit by

\begin{equation}
\tilde{U}\sim \tilde{u}\rho\, .
\end{equation}

For Schwarzschild $p$-branes $\tilde{u}=\omega/2$ and the above formula
gives the knownresult $T_{p}=-\omega/2$.

Finally, let us just stress that the tensions, temperature and entropy density
of the $d$-dimensional black $p$-branes that we are studying coincide with the
mass, temperature and entropy of the $(\tilde{p}+4)$-dimensional black hole
that one finds by toroidal dimensional reduction over the $p$ spacelike
worldvolume directions.


\subsection{JNW black branes}

The Janis-Newman-Winicour (JNW) black branes can be obtained by uplifting the
4-dimensional JNW solutions \cite{Janis:1968zz,Agnese:1985xj} to $d=4+p$
dimensions. The latter are static, spherically-symmetric solutions of the
Einstein-dilaton theory

\begin{equation}
\mathcal{I}[g_{\mu\nu},\varphi]
=
\int d^{4}x \left[ R +2(\partial\varphi)^{2} \right]\, ,  
\end{equation}

\noindent
which depend on two independent parameters: the mass $M$ and the scalar charge
$\Sigma$ defined asymptotically ($r\rightarrow \infty$) by\footnote{We set a
  third possible parameter, which is the asymptotic value of the scalar, to
  zero.}

\begin{equation}
g_{tt}\sim  1-\frac{2M}{r}\, ,
\hspace{1cm}
e^{\varphi} \sim 1+\frac{\Sigma}{r}\, . 
\end{equation}

\noindent
They can be written in the form

\begin{equation}
\label{eq:JNWALCsolutions}
\begin{array}{rcl}
ds^{2} & = & W^{\frac{2M}{\omega}-1}Wdt^{2} 
-W^{1-\frac{2M}{\omega}}\left[ W^{-1}dr^{2} 
+r^{2}d\Omega_{(2)}^{2}\right]\, ,  \\
& & \\
\varphi & = & \frac{\Sigma}{\omega}\log{W}\, ,\\
\end{array}
\end{equation}

\noindent
where the function $W$ is given has the same form as in the 4-dimensional
Schwarzschild black hole ($\tilde{p}=0$) Eq.~(\ref{eq:dSchwar})

\begin{equation}
W  =  1+\frac{\omega}{r}\, ,
\end{equation}

\noindent
and where the integration constant $\omega$ is related to $M$ and $\Sigma$ by

\begin{equation}
\omega  =  - 2\sqrt{M^{2} + \Sigma^{2}}\, .  
\end{equation}

These solutions, which are asymptotically flat, are singular if $\Sigma\neq
0$, in agreement with the no-hair theorem: the area of the 2-spheres vanishes
for $r=\omega$ and there is no regular event horizon. Not only the metric is
singular there: $e^{\varphi}$ also vanishes for $r=-\omega$.  When $\Sigma=0$
the solution reduces to Schwarzschild's.

Using the formulae of Appendix~\ref{sec-dimred} we can uplift these solutions
to solutions of pure $4+p$ gravity with metrics given by

\begin{equation}
ds^{2}_{(4+p)}
=
W^{-\frac{2}{\omega}\left[M+\sqrt{\frac{p}{p+4}} \Sigma\right]}dt^{2}
-W^{\frac{4 \Sigma}{\omega\sqrt{p(p+4)}}}d\vec{y}^{\, 2}_{(p)}
-W^{\frac{2}{\omega}\left[M\sqrt{\frac{p}{p+4}} \Sigma\right]}  
\left[dr^{2} +W r^{2} d\Omega^{2}_{(2)} \right]\, ,
\end{equation}

\noindent
and, in FGK coordinates, by

\begin{equation}
ds^{2}_{(4+p)}
=
e^{-2\left[M+\sqrt{\frac{p}{p+4}} \Sigma\right]\rho}dt^{2}
-e^{\frac{4 \Sigma}{\sqrt{p(p+4)}}\rho}d\vec{y}^{\, 2}_{(p)}
-e^{2\left[M-\sqrt{\frac{p}{p+4}} \Sigma\right]\rho} 
\gamma_{(3)\, \underline{m}\underline{n}}dx^{m}dx^{n}\, ,  
\end{equation}

\noindent
which fits in the general form Eq.~(\ref{eq:generalmetric1}) with

\begin{equation}
e^{-\tilde{U}} =   e^{\left[M+\sqrt{\frac{p}{p+4}} \Sigma\right]\rho}\, ,
\hspace{1cm}
W= e^{-2\left[M+\frac{p+2}{\sqrt{p(p+4)}} \Sigma\right]\rho}\, .
\end{equation}

\noindent
The asymptotic behaviors of $\tilde{U}$ and $W$ are different from those in
Eqs.~(\ref{eq:asympU}) and (\ref{eq:asympW}) and the solution is, therefore,
doubly singular: the areas of the 2-spheres vanish on the horizons and the
worldvolume metric is also singular there.



\begin{thebibliography}{99}

\raggedright

\bibitem{Ferrara:1997tw}
S.~Ferrara, G.~W.~Gibbons, R.~Kallosh,
Nucl.\ Phys.\  {\bf B500 } (1997)  75-93.
[\hepth{9702103}].

\bibitem{Ferrara:1995ih}
S.~Ferrara, R.~Kallosh, A.~Strominger,
Phys.\ Rev.\  {\bf D52 } (1995)  5412-5416.
[\hepth{9508072}].
A.~Strominger,
Phys.\ Lett.\  {\bf B383 } (1996)  39-43.
[\hepth{9602111}].
S.~Ferrara, R.~Kallosh,
Phys.\ Rev.\  {\bf D54 } (1996)  1514-1524.
[\hepth{9602136}].
S.~Ferrara, R.~Kallosh,
Phys.\ Rev.\  {\bf D54 } (1996)  1525-1534.
[\hepth{9603090}].

\bibitem{Meessen:2010fh}
P.~Meessen, T.~Ort\'{\i}n and  S.~Vaul\`a,
JHEP {\bf 11} (2010) 072.
[\arxiv{1006.0239} [hep-th]].

\bibitem{Galli:2011fq}
P.~Galli, T.~Ort\'{\i}n, J.~Perz, C.~S.~Shahbazi,
JHEP {\bf 1107 } (2011)  041.
[\arxiv{1105.3311}].

\bibitem{Mohaupt:2011aa}
T.~Mohaupt and O.~Vaughan,
\arxiv{1112.2876}.

\bibitem{Meessen:2011mu}
P.~Meessen, T.~Ort\'{\i}n, J.~Perz and C.~S.~Shahbazi,
To be published in \textit{Physics Letters}~\textbf{B}
\arxiv{1112.3332}.

\bibitem{Meessen:2011bd}
P.~Meessen, T.~Ort\'{\i}n,
Phys.\ Lett.\ B {\bf 707} (2012) 178.
[\arxiv{1107.5454}].

\bibitem{Mohaupt:2009iq}
T.~Mohaupt and K.~Waite,
JHEP {\bf 0910} (2009) 058
[\arxiv{0906.3451}].

\bibitem{Mohaupt:2010fk}
T.~Mohaupt, O.~Vaughan,
Class.\ Quant.\ Grav.\  {\bf 27 } (2010)  235008.
[\arxiv{1006.3439}].

\bibitem{Ortin:2011vm}
T.~Ort\'{\i}n,
Phys.\ Lett.\ B {\bf 700} (2011) 261
[\arxiv{1103.2738}].

\bibitem{kn:MOPS}
P.~Meessen, T.~Ort\'{\i}n, J.~Perz and C.~S.~Shahbazi,
to appear.

\bibitem{Chemissany:2011gr}
W.~Chemissany, B.~Janssen and T.~Van Riet,
JHEP {\bf 1110} (2011) 002
[\arxiv{1107.1427} [hep-th]].

\bibitem{Horowitz:1991cd}
G.~T.~Horowitz and A.~Strominger,
Nucl.\ Phys.\ B {\bf 360} (1991) 197.


\bibitem{Bergshoeff:2004kh}
E.~Bergshoeff, S.~Cucu, T.~de Wit, J.~Gheerardyn, S.~Vandoren 
and A.~Van Proeyen,
Class.\ Quant.\ Grav.\  {\bf 21} (2004) 3015
[\hepth{0403045}].

\bibitem{Bellorin:2006yr}
J.~Bellor\'{\i}n, P.~Meessen, T.~Ort\'{\i}n,
JHEP {\bf 0701 } (2007)  020.
[\hepth{0610196}].

\bibitem{Gauntlett:2002nw}
J.P.~Gauntlett, J.B.~Gutowski, C.M.~Hull, S.~Pakis and H.S.~Reall,
Class.\ Quant.\ Grav.\  {\bf 20} (2003) 4587
[\hepth{0209114}].
J.B.~Gutowski and H.S.~Reall,
JHEP {\bf 0404} (2004) 048
[\hepth{0401129}].
J.B.~Gutowski and W.~Sabra,
JHEP {\bf 0510} (2005) 039
[\hepth{0505185}].

\bibitem{Ortin:2004ms}
T.~Ort\'{\i}n,
``Gravity and strings,''
Cambridge Unversity, Cambridge University Press, 2004

\bibitem{Janssen:2007rc}
B.~Janssen, P.~Smyth, T.~Van Riet and B.~Vercnocke,
JHEP {\bf 0804} (2008) 007
[\arxiv{0712.2808} [hep-th]].
E.~Bergshoeff, W.~Chemissany, A.~Ploegh, M.~Trigiante and T.~Van Riet,
Nucl.\ Phys.\ B {\bf 812} (2009) 343
[\arxiv{0806.2310} [hep-th]].

\bibitem{Tangherlini:1963bw}
F.~R.~Tangherlini,
Nuovo Cim.\  {\bf 27} (1963) 636.

\bibitem{Agnese:1985xj}
A.~G.~Agnese and M.~La Camera,
Phys.\ Rev.\ D {\bf 31} (1985) 1280.

\bibitem{Janis:1968zz}
A.~I.~Janis, E.~T.~Newman and J.~Winicour,
Phys.\ Rev.\ Lett.\  {\bf 20} (1968) 878.

\bibitem{Myers:1986un}
R.~C.~Myers and M.~J.~Perry,
Annals Phys.\  {\bf 172} (1986) 304.

\bibitem{Myers:1999psa}
R.~C.~Myers,
Phys.\ Rev.\ D {\bf 60} (1999) 046002
[\hepth{9903203}].












\end{thebibliography}
\end{document}